\documentclass[aip,graphicx]{revtex4-2}
\usepackage{graphicx}
\usepackage{amsmath}
\usepackage{amsfonts}
\usepackage{amssymb}

\usepackage{comment}

\usepackage{color} 
\usepackage[normalem]{ulem} 
\usepackage{tikz} 
\usepackage{pgfplots} 
\usepackage{array} 
\usepackage{multirow}  

\usepackage[bookmarksopen=false]{hyperref} 

\newlength{\mm}
\newlength{\cm}
\newcommand{\ie}{\textit{i.e.}}

\newcommand{\mbigskip}{\vspace{1mm}}

\providecommand\bcdot{\boldsymbol{\cdot}}

\newcommand{\ud}{\mathrm{d}}
\newcommand\mnewcommand[1]{%
\let#1\relax \newcommand#1 }

\mnewcommand{\mysubsubsection}{\subsubsection}
\mnewcommand{\comentar}[1]{}

\mnewcommand{\Nx}{N_{\x}}
\mnewcommand{\Ny}{N_{\y}}
\mnewcommand{\Nz}{N_{\z}}
\mnewcommand{\Nm}{N_m}
\mnewcommand{\Ns}{N_s}
\mnewcommand{\complconj}[1]{#1^{*}}
\mnewcommand{\mvbrack}[1]{\left[ #1 \right]}
\mnewcommand{\step}{\Delta}
\mnewcommand{\dt}{\step t}
\mnewcommand{\traspose}{^{T}}
\mnewcommand{\avgtime}[1]{\left< #1 \right>}
\mnewcommand{\avg}[1]{\overline{#1}}
\mnewcommand{\mcdot}{\bcdot}
\mnewcommand{\mnabla}{\nabla}

\mnewcommand{\real}{\mathbb{R}}
\mnewcommand{\complex}{\mathbb{C}}
\mnewcommand{\imag}{i\real}

\mnewcommand{\realpart}{\mathsf{Re}}
\mnewcommand{\imagpart}{\mathsf{Im}}

\mnewcommand{\foutrans}[1]{\hat{#1}}
\mnewcommand{\modetrans}[1]{\tilde{#1}}

\mnewcommand{\timeintparam}{\kappa}

\mnewcommand{\x}{x}
\mnewcommand{\y}{y}
\mnewcommand{\z}{z}

\newcommand{\xvec}{\isvector{x}}

\mnewcommand{\vels}{u_s}
\mnewcommand{\uvel}{u_1}
\mnewcommand{\vvel}{u_2}
\mnewcommand{\wvel}{u_3}

\mnewcommand{\facevel}{[\velh]_\nface}

\mnewcommand{\flux}{f}
\mnewcommand{\Dx}{\Delta x}
\mnewcommand{\Dy}{\Delta \y}

\mnewcommand{\isvector}[1]{\boldsymbol{#1}}
\mnewcommand{\istensor}[1]{\mathsf{#1}}

\mnewcommand{\va}{\isvector{a}}
\mnewcommand{\vb}{\isvector{b}}
\mnewcommand{\vc}{\isvector{c}}
\mnewcommand{\vd}{\isvector{d}}
\mnewcommand{\sca}{\phi}
\mnewcommand{\scafield}{\isvector{\sca}}
\mnewcommand{\scafieldc}{\scafield_{c}}

\mnewcommand{\vel}{\isvector{u}}
\mnewcommand{\velv}{\isvector{v}}
\mnewcommand{\velw}{\isvector{w}}
\mnewcommand{\velz}{\isvector{z}}
\mnewcommand{\normal}{\isvector{n}}

\mnewcommand{\veluc}{\isvector{u}_1}
\mnewcommand{\velvc}{\isvector{u}_2}
\mnewcommand{\velwc}{\isvector{u}_3}

\mnewcommand{\basis}{\isvector{w}}

\mnewcommand{\tensor}{\istensor{T}}
\mnewcommand{\Identity}{\istensor{I}}

\mnewcommand{\bodyforce}{\isvector{f}}
\mnewcommand{\velh}{\vel_s}
\mnewcommand{\velhc}{\vel_c}
\mnewcommand{\velhcCB}[1]{\ifthenelse{\equal{#1}{}}{\velhc^{\ominus}}{\velhc^{#1,\ominus}}}
\mnewcommand{\velhcCBFree}[1]{\ifthenelse{\equal{#1}{}}{\velhc^{\oplus}}{\velhc^{#1,\oplus}}}
\mnewcommand{\velhcHF}[1]{\ifthenelse{\equal{#1}{}}{\velhc^{>}}{\velhc^{#1,>}}}
\mnewcommand{\velhcLF}[1]{\ifthenelse{\equal{#1}{}}{\velhc^{<}}{\velhc^{#1,<}}}
\mnewcommand{\velhsHF}[1]{\ifthenelse{\equal{#1}{}}{\velh^{>}}{\velh^{#1,>}}}
\mnewcommand{\velhsLF}[1]{\ifthenelse{\equal{#1}{}}{\velh^{<}}{\velh^{#1,<}}}
\mnewcommand{\vortd}{\vort_v}
\mnewcommand{\presh}{\isvector{p}_c}
\mnewcommand{\pseudopresh}{{\isvector{\tilde{p}}}_c}
\mnewcommand{\bodyforceh}{\bodyforce_c}
\mnewcommand{\diver}{\mnabla \cdot}
\mnewcommand{\lapl}{\mnabla^2}
\mnewcommand{\grad}{\mnabla}
\mnewcommand{\Ru}[1]{\isvector{R} \left( #1 \right)}
\mnewcommand{\Rud}[1]{\ifthenelse{\equal{#1}{}}{\mathsfbi{R}}{\istensor{R} \left( #1 \right)}}
\mnewcommand{\vecnull}{\isvector{0}}
\mnewcommand{\vecnulls}{\vecnull_{s}}
\mnewcommand{\vecnullg}{\vecnull_{h}}
\mnewcommand{\vecnullc}{\vecnull_{c}}
\mnewcommand{\vecone}{\isvector{1}}
\mnewcommand{\veconec}{\vecone_{c}}
\mnewcommand{\vecones}{\vecone_{s}}
\mnewcommand{\veconeg}{\vecone_{h}}
\mnewcommand{\veconetresc}{\vecone_{3c}}

\mnewcommand{\velhg}{\vel_h}
\mnewcommand{\preshg}{\isvector{p}_h}

\mnewcommand{\dim}{3}
\mnewcommand{\Ndim}{d}

\mnewcommand{\ud}{d}
\mnewcommand{\vort}{\mathbf{w}}
\mnewcommand{\rot}[1]{\mnabla \times #1}
\mnewcommand{\selfinnerprod}[1]{\innerprod{#1}{#1}}
\mnewcommand{\innerprod}[2]{< #1 , #2 >}
\mnewcommand{\convective}[2]{C \left( #1 , #2 \right)}
\mnewcommand{\intvol}[1]{\int_{\Omega} #1 \ud \Omega}
\mnewcommand{\intsurf}[1]{\int_{\partial \Omega} #1 \ud S}

\mnewcommand{\nvc}{k}
\mnewcommand{\nedge}{v}
\mnewcommand{\axis}{i}
\mnewcommand{\nface}{f}
\mnewcommand{\cp}{c}
\mnewcommand{\cpA}{{c1}}
\mnewcommand{\cpB}{{c2}}
\mnewcommand{\Fedge}[1]{F_e ( #1 )}
\mnewcommand{\Fcell}[1]{F_f ( #1 )}
\mnewcommand{\Fvolume}[1]{F_c ( #1 )}

\mnewcommand{\mathsfbi}[1]{\mathsf{#1}}

\mnewcommand{\conv}{\mathsfbi{C}\left( \velh \right)}
\mnewcommand{\convg}{\mathsfbi{C}\left( \velhg \right)}
\mnewcommand{\convc}{\mathsfbi{C}_{c} \left( \velh \right)}
\mnewcommand{\convu}{\mathsfbi{C}_{u} \left( \velh \right)}
\mnewcommand{\convtraspose}{\mathsfbi{C}\traspose\left( \velh \right)}
\mnewcommand{\convctraspose}{\mathsfbi{C}_{c}\traspose\left( \velh \right)}
\mnewcommand{\convgtraspose}{\mathsfbi{C}\traspose\left( \velhg \right)}
\mnewcommand{\convarg}[1]{\mathsfbi{C}\left( #1 \right)}
\mnewcommand{\convcarg}[1]{\mathsfbi{C}_{c}\left( #1 \right)}
\mnewcommand{\convargtraspose}[1]{\mathsfbi{C}\traspose\left( #1 \right)}
\mnewcommand{\convutraspose}{\mathsfbi{C}_{u} \traspose\left( \velh \right)}
\mnewcommand{\convmat}{\mathsfbi{C}}
\mnewcommand{\convUP}{\convmat^{\UP}_{c} \left( \velh \right)}

\DeclareRobustCommand{\spreg}[2]{\ifthenelse{\equal{#2}{}}{\convmat_{#1}}{\convmat_{#1}\left( #2 \right)}}
\mnewcommand{\velauxc}{\velv_c}
\mnewcommand{\velauxWc}{\velw_c}
\mnewcommand{\filter}{\mathsfbi{G}_\epsilon}

\mnewcommand{\velhcmode}[1]{\ifthenelse{\equal{#1}{+}}{\velhcHF{}}{\ifthenelse{\equal{#1}{-}}{\velhcLF{}}{\ifthenelse{\equal{#1}{=}}{\velhc}{}}}}
\mnewcommand{\velhsmode}[1]{\ifthenelse{\equal{#1}{+}}{\velhsHF{}}{\ifthenelse{\equal{#1}{-}}{\velhsLF{}}{\ifthenelse{\equal{#1}{=}}{\velh}{}}}}

\mnewcommand{\velhctype}[2]{\ifthenelse{\equal{#2}{f}}{\overline{\velhcmode{#1}}}{\ifthenelse{\equal{#2}{r}}{\left(\velhcmode{#1}\right)^\prime}{\ifthenelse{\equal{#2}{v}}{\velhcmode{#1}}{}}}}
\mnewcommand{\velhstype}[2]{\ifthenelse{\equal{#2}{f}}{\overline{\velhsmode{#1}}}{\ifthenelse{\equal{#2}{r}}{\left(\velhsmode{#1}\right)^\prime}{\ifthenelse{\equal{#2}{v}}{\velhsmode{#1}}{}}}}

\mnewcommand{\triadic}[6]{\left(\velhctype{#1}{#2}\right)\traspose \convarg{\velhstype{#3}{#4}} \velhctype{#5}{#6}}

\mnewcommand{\diff}{\mathsfbi{D}}
\mnewcommand{\diffg}{\mathsfbi{D}}
\mnewcommand{\diffc}{\diff_{\cp}}
\mnewcommand{\diffu}{\diff_{u}}
\mnewcommand{\dive}{\mathsfbi{M}}
\mnewcommand{\divec}{\mathsfbi{M}_c}
\mnewcommand{\diveg}{\mathsfbi{M}_h}
\mnewcommand{\divescaf}{\dive_{sf}}
\mnewcommand{\divevecf}{\dive_{vf}}
\mnewcommand{\graddc}{\mathsfbi{G}_c}
\mnewcommand{\gradd}{\mathsfbi{G}}
\mnewcommand{\graddg}{\mathsfbi{G}_h}
\mnewcommand{\lapld}{\mathsfbi{L}}
\mnewcommand{\lapldc}{\lapld_{c}}
\mnewcommand{\vcvects}{\mathsfbi{\Omega}_s}
\mnewcommand{\pseudovcvects}{\tilde{\mathsfbi{\Omega}}_s}
\mnewcommand{\vcvectg}{\mathsfbi{\Omega}_h}
\mnewcommand{\vcvectc}{\mathsfbi{\Omega}_c}
\mnewcommand{\vcvectv}{\mathsfbi{\Omega}_v}
\mnewcommand{\vcvectvc}{\mathsfbi{\Omega}}
\mnewcommand{\vcvect}{\mathsfbi{\Omega}}
\mnewcommand{\nullmat}{\mathsfbi{0}}
\mnewcommand{\normd}[1]{|| #1 ||}
\mnewcommand{\rotd}{\mathsfbi{R}}
\mnewcommand{\graddscaf}{\gradd_{sf}}
\mnewcommand{\graddvecf}{\gradd_{vf}}
\mnewcommand{\graddx}{\gradd_{\x}}
\mnewcommand{\graddy}{\gradd_{\y}}
\mnewcommand{\graddz}{\gradd_{\z}}
\mnewcommand{\graddd}[1]{\ifthenelse{\equal{#1}{1}}{\graddx}{\ifthenelse{\equal{#1}{2}}{\graddy}{\ifthenelse{\equal{#1}{3}}{\graddz}{\gradd_{x_i}}}}}
\mnewcommand{\graddprod}[2]{\graddd{#1}\traspose \vcvect \graddd{#2}}

\mnewcommand{\fluxh}[2]{T_{#1}\left({#2}\right)}

\mnewcommand{\twoD}{two-dimensional}
\mnewcommand{\threeD}{three-dimensional}
\mnewcommand{\TwoD}{Two-dimensional}
\mnewcommand{\ThreeD}{Three-dimensional}
\mnewcommand{\biD}{\twoD~}
\mnewcommand{\triD}{\threeD~}
\mnewcommand{\BiD}{\TwoD~}
\mnewcommand{\TriD}{\ThreeD~}
\mnewcommand{\biandtriD}{two- and \threeD~}
\mnewcommand{\BiandtriD}{Two- and \threeD~}
\mnewcommand{\Dim}[1]{\ifthenelse{\equal{#1}{2}}{\twoD}{\ifthenelse{\equal{#1}{3}}{\threeD}{KK}}}

\mnewcommand{\inttypeoftext}{\mathsf}
\mnewcommand{\kinener}{\inttypeoftext{E}}
\mnewcommand{\enstrophy}{\inttypeoftext{\Omega}}
\mnewcommand{\helicity}{\inttypeoftext{H}}
\mnewcommand{\vorthelicity}{\helicity_{\vort}}
\mnewcommand{\palinstrophy}{\inttypeoftext{P}}

\mnewcommand{\helicityd}{\helicity_c}
\mnewcommand{\enstrophyd}{\enstrophy_c}

\mnewcommand{\vvlength}{m}
\mnewcommand{\pvlength}{n}
\mnewcommand{\evlength}{e}

\mnewcommand{\Isc}{\Gamma}
\mnewcommand{\Ics}{\Isc\traspose}
 
\mnewcommand{\Scs}{\Gamma_{c \rightarrow s}}
\mnewcommand{\Ssc}{\Gamma_{s \rightarrow c}}

\mnewcommand{\Sscal}{\Pi_{c \rightarrow s}}
\mnewcommand{\Sthreescal}{\Pi}
\mnewcommand{\NormalVect}[1]{\ifthenelse{\equal{#1}{}}{\istensor{N}_{s}}{\istensor{N}_{s,#1}}}

\mnewcommand{\Correction}{\istensor{P}}
\mnewcommand{\PseudoCorrection}{\tilde{\Correction}}
\mnewcommand{\kernel}[1]{Ker \left( #1 \right)}
\mnewcommand{\sCorrection}{\Correction_s}
\mnewcommand{\cCorrection}{\PseudoCorrection_c}

\mnewcommand{\Ivs}{\Psi}
\mnewcommand{\Isv}{\Psi\traspose}

\mnewcommand{\OIsc}{\Upsilon}
\mnewcommand{\OIcs}{\Pi}

\mnewcommand{\order}{o}
\mnewcommand{\coarsemesh}{i}
\mnewcommand{\sizevc}{{V}}
\mnewcommand{\error}{\epsilon}

\mnewcommand{\convH}{\text{(Conv)}_{\helicityd}}
\mnewcommand{\diffH}{\text{(Diff)}_{\helicityd}}
\mnewcommand{\presH}{\text{(Pres)}_{\helicityd}}

\newcommand{\Dt}{\dt}

\newcommand{\UP}{\mathrm{UP}}

\newcommand{\press}{p}

\mnewcommand{\vapf}{\lambda}
\mnewcommand{\angle}{\varphi}
\mnewcommand{\vapfn}{-e^{-i\angle}}
\mnewcommand{\dtn}{\widetilde{\dt}}
\mnewcommand{\vapg}{\lambda^{\pm}}
\mnewcommand{\Kopt}{K_{opt}}
\mnewcommand{\Topt}{T_{opt}}
\mnewcommand{\CFLAB}{CFL+AB2}
\mnewcommand{\NewMeth}{AlgEigCD+\timeintparam 1L2}
\mnewcommand{\AlgEigCD}{{\it AlgEigCD}}
\mnewcommand{\EigenCD}{{\it EigenCD}}
\mnewcommand{\parNiki}{\ULL{q}}

\mnewcommand{\scalResK}{\beta}
\mnewcommand{\ResConst}{C_{r}}
\mnewcommand{\scalConstRey}{\kappa}
\mnewcommand{\scalKmaxRey}{\gamma}
\mnewcommand{\KmaxConst}{C_{k}}
\mnewcommand{\scalResReyPhy}{\tilde{\beta}}

\mnewcommand{\scalResDt}{q} 
\mnewcommand{\scalDtRey}{\alpha}
\mnewcommand{\scalResReyNum}{\tilde{\alpha}}
\mnewcommand{\scalnRey}{\xi}

\newcommand{\pSLa}{S1}
\newcommand{\pSLb}{S2}
\newcommand{\pSLc}{S3}
\newcommand{\pSLd}{S4}
\newcommand{\pSLe}{S5}
\newcommand{\pWa}{W1}
\newcommand{\pWb}{W2}
\newcommand{\pWc}{W3}
\newcommand{\pWd}{W4}

\newcommand{\pPLa}{P1}
\newcommand{\pPLb}{P2}
\newcommand{\pPLc}{P3}
\newcommand{\pPLd}{P4}
\newcommand{\pPLe}{P5}

\newcommand{\betaf}[2]{\mathcal{B} ( #1 , #2 )}
\newcommand{\incbetaf}[3]{\mathcal{B}_{#1} ( #2 , #3 )}
\newcommand{\half}{1/2}

\newcommand{\Gten}{\istensor{G}}
\newcommand{\QG}{Q_{\Gten}}
\newcommand{\RG}{R_{\Gten}}

\newcommand{\aft}[1]{#1}
\newcommand{\aftR}[1]{#1}
\newcommand{\bef}[1]{}

\newcommand{\ULL}[1]{\textcolor{red}{\bf #1}}

\definecolor{darkgreen}{rgb}{0.0,0.5,0.0}

\newcommand{\Xavi}[1]{#1}
\newcommand{\XaviR}[1]{#1}
\newcommand{\Adel}[1]{#1} 

\begin{document}


\title{On the Reynolds-number scaling of Poisson solver complexity}




\newcommand{\cttc}{\affiliation{Heat and Mass Transfer Technological Center, Technical University of Catalonia,\\
c/Colom 11, 08222 Terrassa (Barcelona), Spain}}

\newcommand{\udp}{\affiliation{Department ICEA, University of Padova, Via Francesco Marzolo, 9, 35131 Padova PD, Italy}}

\author{F.X.Trias}
\email[]{francesc.xavier.trias@upc.edu}
\cttc

\author{A.Alsalti-Baldellou}
\email[]{adel.alsaltibaldellou@unipd.it}
\cttc
\udp

\author{A.Oliva}
\email[]{asensio.oliva@upc.edu}
\cttc


\date{\today}


\begin{abstract}
  We aim to answer the following question: {\it is the complexity of
    numerically solving Poisson’s equation increasing or decreasing
    for very large simulations of incompressible flows?} Physical and
  numerical arguments are combined to derive power-law scalings at
  very high Reynolds numbers. A theoretical convergence analysis for
  both Jacobi and multigrid solvers defines a two-dimensional phase
  space divided into two regions depending on whether the number of
  solver iterations tends to decrease or increase with the Reynolds
  number. Numerical results indicate that, for \XaviR{Navier--Stokes}
  turbulence, the complexity decreases with increasing Reynolds
  number, whereas for the one-dimensional Burgers’ equation it follows
  the opposite trend. The proposed theoretical framework thus provides
  a unified perspective on how solver \Xavi{convergence} scales with
  $Re$-number and offers valuable guidance for the development of
  next-generation preconditioning and multigrid strategies for
  extreme-scale simulations.
\end{abstract}



\maketitle 

\section{Introduction}

\label{intr}

We consider the direct numerical simulation (DNS) of turbulent
incompressible flows. \Xavi{For clarity in the forthcoming analysis,}
we restrict attention to Newtonian fluids with constant physical
properties. \Xavi{This assumption does not entail any loss of
  generality for the arguments developed here.} Under these
assumptions, the governing \Xavi{Navier--Stokes} (NS) equations in
non-dimensional form read
\begin{equation}
\label{NS_eqs}
\partial_t \vel + ( \vel \cdot \nabla ) \vel = \frac{1}{Re} \lapl \vel - \grad \press, \hspace{6.69mm} \diver \vel = 0 ,
\end{equation}
\noindent where $\vel(\xvec,t)$ and $p(\xvec,t)$ denote the velocity
and pressure fields, respectively, and $Re = U l / \nu$ is the
Reynolds number. Here, $\nu$ is the kinematic viscosity, while $U$ and
$l$ denote the characteristic velocity and length scale, respectively,
which are typically associated with the motion of the largest
\Xavi{flow} scales.

\mbigskip

Then, these equations must be discretized in both space and time. For
the spatial discretization, a wide variety of numerical methods and
schemes are available~\cite{FER20}. \Xavi{Their} choice depends on
factors such as local accuracy, numerical stability, boundedness, and
the conservation of global quantities such as momentum and kinetic
energy, among others. The earliest DNS studies of turbulent flows were
restricted to simple configurations, primarily homogeneous isotropic
turbulence (HIT) simulations~\cite{ORZ72,PUL95,CAO99} and turbulent
channel flows~\cite{KIM87,MOS99} at moderate $Re$-numbers. These
simulations relied on Fourier (or Fourier–Chebyshev for channel flows)
pseudospectral methods, combined with dealiasing techniques to treat
the nonlinear convective terms~\cite{CAN88}. Over the past decades,
advances in numerical algorithms and high-performance computing (HPC)
systems enabled DNS at higher
$Re$-numbers~\cite{KAN04,HOY06,ISH09,STEV11,LEE15,DABTRI19-3DTOPO-RB,PIR25}
and more complex \XaviR{flow
  configurations\cite{VIN15,TRI14-CYL,PONTRI18-BFS,KER24,PAR24,DUO24,LIN25,GAO25,KOU25}}. In
parallel, community-accessible resources such as the Johns Hopkins
Turbulence Database~\cite{PER07-JHTDB,LI08-JHTDB} have provided
researchers with unprecedented access to large-scale DNS datasets,
including homogeneous isotropic turbulence up to $Re_\lambda \approx
2500$ at $32768^3$ resolution, as well as other canonical flows. While
Fourier-based methods remain the standard for canonical
configurations, mesh-based approaches such as finite-volume,
finite-difference, and finite-element methods have become essential to
simulate turbulence in complex geometries. From a physical
perspective, turbulence arises from the intricate interplay between
nonlinear convection, which transfers kinetic energy from large to
small scales, and viscous dissipation, which ultimately balances this
transfer. Numerically, schemes that introduce artificial dissipation
can severely distort this balance at the smallest
scales. Consequently, \Xavi{it is widely accepted within the DNS
  community that}, regardless of the discretization method, reliable
simulations require numerical methods that are virtually free from
artificial
\XaviR{dissipation~\cite{PER11,KOR14,TRI08-JCP,KOM17,VALTRI19-EPLS,ZHA22,COP23}}.

\mbigskip

In addition to spatial discretization, the governing equations must be
advanced in time, which requires addressing both the integration of
the momentum equations and the pressure–velocity coupling inherent to
incompressible flows. Starting from the pioneering simulations of HIT
and channel flows, most DNS simulations of incompressible turbulence
have been performed using fractional-step projection
methods~\cite{CHO68} combined with either explicit or semi-implicit
time integration for the momentum equation. In virtually all cases,
the nonlinear convective term is treated explicitly, which severely
restricts the allowable time-step. Specifically, the eigenvalues of
the linearized system, scaled by the chosen time-step, must remain
within the stability region of the temporal
scheme~\cite{TRI22-AlgEigCD}. This restriction is usually expressed
through the CFL condition~\cite{CFL28}. Consequently, each time-step
requires the solution of a pressure Poisson equation, which usually
represents the dominant computational cost and the main bottleneck in
large-scale DNS \XaviR{of incompressible flows}.

\mbigskip

\aft{In canonical configurations with periodic boundary conditions and
  uniform meshes, this Poisson problem can be solved very efficiently
  using FFT-based direct solvers, which exploit the analytical
  eigenstructure of the discrete Laplacian to obtain the solution to
  machine accuracy~\cite{GORTRI07-CF,DOD14,COS18}. However, such
  approaches are restricted to a limited set of idealized
  configurations~\cite{TRI06-JFM,TRI14-CYL,DABTRI15-TOPO-RB}. Most
  practical simulations involve complex geometries, non-periodic
  boundary conditions, or non-uniform grids, for which FFT-based
  methods are not applicable and general iterative solvers are
  required.} \Adel{\aft{In this regard, one} of the most efficient
  approaches to solve such a Poisson equation is through iterative
  methods based on Krylov subspaces \cite{Saa03}, whose implementation
  is simple and easily parallelisable, requiring only basic linear
  algebra operations.  However, iterative linear solvers must be
  properly preconditioned to be effective \cite{Ben02}.  In this
  sense, while preconditioners based on incomplete factorizations were
  very popular in the early days of numerical linear
  algebra~\cite{MeiVdV77}, their sequential nature and the increasing
  availability of parallel computers made them lose ground against
  alternatives with higher degrees of parallelism, such as sparse
  approximate inverses~\cite{KolYer93,IsoJanBer22}, whose application
  solely relies on the sparse-matrix vector product.}

\mbigskip

\Adel{Regardless of their specific features, none of these methods are
  {\em optimal} in the sense that when augmenting the mesh resolution
  (hence increasing the linear system size), the problem becomes more
  ill-conditioned and more iterations are required to reach the same
  accuracy. This problem worsens nowadays, as cutting-edge DNS require
  solving extreme-scale linear systems on massively parallel
  supercomputers, and single-level preconditioners generally require
  excessive iterations.}

\mbigskip

\Adel{The problem of weak scalability is overcome with multilevel
  preconditioners like geometric or algebraic
  multigrid~\cite{Stuben2001,XuZik17} \XaviR{(MG)}. They combine the
  ``smoothing'' properties of single-level methods with the robustness
  of direct solvers by assembling a hierarchy of grids and taking
  advantage of the fact that smooth error becomes less smooth after
  coarsening.  Then, a single-level preconditioner smooths the error
  at each level, and a direct solver removes the remaining
  low-frequency modes at the coarsest level.  To ensure an effective
  interplay between smoother and coarse-grid correction, the transfer
  operators used to jump between levels (restriction and prolongation)
  must preserve the near-null space of the coefficient matrix.  When
  done accurately, \XaviR{MG} preconditioners provide convergence
  rates independent (or mildly dependent) of the grid size and, owing
  to their parallel efficiency, often exhibit an almost ideal weak
  scaling.}

In summary, reliable numerical techniques for DNS of incompressible
flows in complex geometries are well established. This includes both
advanced spatial \XaviR{discretizations, accurate time-integration
  methods and advanced Poisson solvers}. Nevertheless, the achievable
$Re$-numbers remain constrained by the computational capacity of
modern HPC systems. With the continuous growth of computational power,
it is reasonable to anticipate DNS at progressively higher $Re$ in the
coming decades. This raises a fundamental question: {\it as the
  $Re$-number increases, will the relative cost of solving the
  pressure Poisson equation decrease, remain constant, or instead
  become an even more critical bottleneck?} \aft{In this work, this
  question is addressed in terms of the algorithmic complexity of the
  Poisson solver, measured through the number of iterations required
  to reach convergence. Other factors affecting the overall runtime of
  large-scale simulations, such as communication overhead or
  hardware-dependent effects, are outside the scope of the present
  analysis.}

\mbigskip

\aft{Despite the central role played by the pressure Poisson equation
  in projection-based incompressible flow solvers, the question of how
  its complexity scales with the Reynolds number has received little
  explicit attention. This issue lies at the crossroads of three areas
  that are usually treated separately in the literature. On the one
  hand, classical turbulence theory provides predictions for the
  scaling of the pressure field and its spectral content at high
  Reynolds numbers~\cite{BAT51,PUL95,CAO99,GOT99,GOT01,ZHA16,XU20}. On
  the other hand, the convergence behaviour of iterative solvers for
  elliptic problems, including Jacobi and multigrid methods, has been
  extensively analysed in the numerical analysis
  community~\cite{Stuben2001,XuZik17}. Finally, projection-based
  formulations of the incompressible Navier--Stokes
  equations~\cite{CHO68}, in which a Poisson equation must be solved
  at each time step to enforce the divergence-free constraint,
  constitute the standard approach in DNS and LES of turbulent
  flows. In practice, these methods rely on fractional-step procedures
  in which the convective term is treated explicitly in time, leading
  to a CFL-type restriction on the time step. Consequently, a very
  large number of time steps is required in high-Reynolds number
  simulations, and the pressure Poisson equation must be solved
  repeatedly throughout the computation. While each of these topics
  has been studied extensively on its own, their combined implications
  for the Reynolds-number scaling of the pressure Poisson solver have
  not been explicitly addressed. The present work aims to bridge this
  gap by combining turbulence-scaling arguments with classical
  convergence estimates for iterative solvers and validating the
  resulting predictions through numerical experiments.}

\mbigskip

To answer this \bef{question}\aft{problem}, both physical and
numerical arguments are combined in the next sections. Firstly, we
analyze the spectral distribution of the Poisson solver residual. We
identify the two main competing effects and how the spectral
distribution of the residual scales with the $Re$-number.Then, in
Section~\ref{solver}, we use these findings to study whether the
number of iterations \bef{inside the Poisson's solver increases or
  decreases with $Re$.}  \aft{required by iterative Poisson solvers
  (in particular multigrid-based methods) increases or decreases with
  $Re$.} The theoretical predictions are validated in
Section~\ref{results} through numerical experiments for both the
incompressible \Xavi{NS equations} and the 1D Burgers’ equation. Test
cases for the NS equations include HIT, air-filled
Rayleigh--B\'{e}nard convection at different Rayleigh numbers, and
flow around a square cylinder at different \Xavi{$Re$-numbers}. On the
other hand, the Burgers' equation tests cover a very wide range of
$Re$ allowing a verification of the proposed scaling laws. Finally,
relevant results are summarized and conclusions are given.

\begin{figure}[!t]
  \begin{center}
    \includegraphics[width=0.69\textwidth]{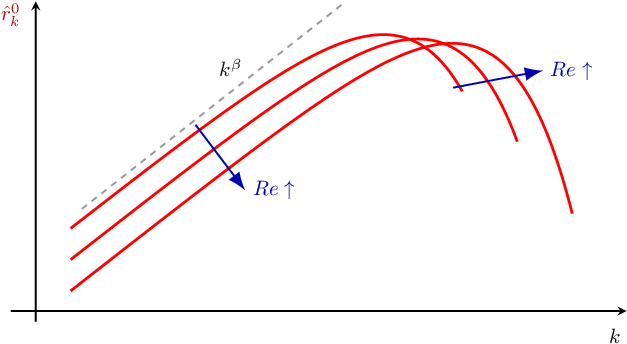}
  \end{center}
\vspace{-3mm}
\caption{Illustrative explanation of the two competing effects on the
  solution of Poisson's equation when increasing $Re$ number:
  time-step, $\Dt$, decreases leading to smaller values of the initial
  residual, $\hat{r}_{k}^{0}$, whereas the range of scales increases.}
\label{Competing_effects}
\end{figure}

\section{ANALYSIS OF THE RESIDUAL OF POISSON'S EQUATION}

\label{residual}


\subsection{Two competing effects}

The steadily increasing capacity of modern HPC systems enables DNS at
higher and higher Reynolds numbers, $Re=Ul/\nu$, where $U$ and $l$
denote the characteristic velocity and length scale of the largest
flow structures. The computational requirements, in terms of the
number of grid points in each direction, $N_{x}$, and time-steps,
$N_{t}$, can be estimated from the classical Kolmogorov
theory~\cite{FRI95} (K41):
\begin{align}
\label{K41_Nx} N^{K41}_{x} &= \frac{L_x}{\Dx} \sim \frac{l}{\eta} \sim Re^{3/4} , \\
\label{K41_Nt} N^{K41}_{t} &= \frac{t_{\text{sim}}}{\Dt} \sim \frac{t_l}{t_{\eta}} \sim \frac{l}{\eta} \frac{u}{U} \sim Re^{3/4} Re^{-1/4} = Re^{1/2} ,
\end{align}
\noindent where $L_x$ and $t_{\text{sim}}$ are the domain size and
total simulation time, assumed to scale with the largest turbulent
structures, \XaviR{\ie~$L_x \sim l$} and $t_{\text{sim}} \sim t_l$,
with $t_{l} \sim l/U$. For DNS resolution, one requires $\Delta x \sim
\eta$ and $\Delta t \sim t_{\eta} \sim \eta/u$, where $\eta$ and $u$
denote the Kolmogorov length and velocity scales, respectively.

\mbigskip

Applying the CFL stability constraints~\cite{TRI22-AlgEigCD,CFL28},
\begin{equation}
\label{CFL}
\Dt^{\text{conv}} \sim \frac{\Dx}{U} \hphantom{kkkkk} \Dt^{\text{diff}} \sim \frac{\Dx^2}{\nu} ,
\end{equation}
\noindent to both the convective and diffusive terms, leads to the
following estimation for the number of time-steps,
\begin{align}
\label{CFL_based_conv} N^{\text{conv}}_{t} \sim \frac{t_l}{\Dt^{\text{conv}}} \sim \frac{l}{U} \frac{U}{l Re^{-3/4}} = Re^{3/4} , \\
\label{CFL_based_diff} N^{\text{diff}}_{t} \sim \frac{t_l}{\Dt^{\text{diff}}} \sim \frac{l}{U} \frac{\nu}{l^2 (Re^{-3/4})^2} = Re^{1/2} .
\end{align}
\noindent Hence, the normalized time-step scales as
\begin{equation}
\label{Dt_scaling}
\frac{\Dt}{t_l} \sim \frac{1}{N_t} \sim Re^{\scalDtRey} \hphantom{kk} \text{with} \hphantom{kk}  \scalDtRey = \left\{ \begin{array}{ll} -1/2 & \text{for K41 (Eq.\ref{K41_Nt}) or diffusion dominant (Eq.\ref{CFL_based_diff})} \\ -3/4 & \text{for convection dominant as Eq.(\ref{CFL_based_conv})} \end{array} \right. 
\end{equation}
%

\mbigskip

\noindent In summary, increasing $Re$ simultaneously requires (i)
larger computational grids and (ii) smaller time-steps. These two
effects act in opposite directions regarding the convergence of the
pressure Poisson equation: larger meshes increase the condition number
of the discrete \XaviR{Laplacian} operator, while smaller time-steps
improve the quality of the initial guess. The central question is
thus: {\it which effect dominates at very high $Re$?}


\subsection{Reynolds number scaling of the solver residual}

\label{Re_scalings}

Although FFT-based direct solvers are very efficient for canonical
flows with periodic directions~\cite{GORTRI07-CF,DOD14,COS18}, for
extreme-scale simulations in complex geometries \XaviR{MG methods} are
expected to be the method of choice. \Adel{Their fast convergence
  results from the complementary roles played by the smoother, which
  is responsible for damping high-frequency error components, and the
  coarse-grid correction, which in turn reduces low-frequency modes.}
We therefore analyze the residual of the Poisson equation as a
function of the $Re$-number, focusing on two key aspects: its
magnitude and its spectral distribution. To study them, we consider a
fractional step method~\cite{CHO68} where $\vel^{p}$ is the predictor
velocity. Imposing that $\diver \vel^{n+1} = 0$, leads to a Poisson
equation for the pressure \XaviR{field}, $\press^{n+1}$,
\begin{equation}
\label{Poisson}
\vel^{n+1} = \vel^{p} - \Dt \grad \press^{n+1} \hphantom{kkk} \stackrel{\diver}{\Longrightarrow} \hphantom{kkk} \lapl \press^{n+1} = \XaviR{\frac{1}{\Dt}} \diver \vel^{p} .
\end{equation}
\Xavi{Assuming a constant time step $\Delta t$,} and taking $p^n$ as
the initial guess, the initial residual becomes
\begin{equation}
\label{residual1}
r^0 = \lapl \press^{n} - \frac{1}{\Dt} \diver \vel^{p,n+1} \stackrel{(\ref{Poisson})}{=} \XaviR{\frac{1}{\Dt}} \left( \diver \vel^{p,n} - \diver \vel^{p,n+1} \right) \approx \partial_t \diver \vel^p \Xavi{,}
\end{equation}
\noindent \Xavi{where $\vel^{p,n}$ and $\vel^{p,n+1}$ denote the
  predictor velocities used to compute $\vel^{n}$ and $\vel^{n+1}$,
  respectively. Note that we assume that the incompressibility
  constraint is satisfied at the previous time-step, \ie~$\diver
  \vel^{n} = 0$.} \XaviR{In practice, this condition is satisfied only
  up to the user-prescribed residual of the Poisson solver, \ie~$|
  \diver \vel^{n} | \le \epsilon$. Nevertheless, the following
  analysis remains valid provided that the absolute variation of the
  initial residual between consecutive time steps is much larger than
  this tolerance, \ie~$|r^{0,n+1} - r^{0,n}| \gg \epsilon$.}

\mbigskip

Alternatively, we can also consider $\tilde{r}^0 = \Dt r^0$. In this
case, the residual reads
\begin{equation}
\label{residual2}
\tilde{r}^0 = \lapl \tilde{\press}^{n} - \diver \vel^{p,n+1} \stackrel{(\ref{Poisson})}{=} \left( \diver \vel^{p,n} - \diver \vel^{p,n+1} \right) \approx \Dt \partial_t \diver \vel^p ,
\end{equation}
\noindent where $\tilde{\press} = \press \Dt$ is a
\XaviR{pseudo-pressure, \ie~pressure re-scaled by $\Dt$}. Notice that
the scaled residual, $\tilde{r}^0$, is physically more meaningful, as
it directly measures the accuracy with which the incompressibility
constraint is imposed. \aft{In projection-based incompressible
  solvers, the purpose of the Poisson equation is precisely to enforce
  $\diver \vel = 0$. Therefore, the residual of the Poisson equation
  quantifies the remaining violation of the divergence-free
  condition. Although the algebraic error, $\isvector{e}$, and the
  residual, $\isvector{r}$, are related through $\isvector{r} = \lapld
  \isvector{e}$, where $\lapld$ is the discrete Laplacian operator,
  the residual provides a direct measure of how accurately the
  incompressibility constraint is satisfied and is the quantity used
  in practical stopping criteria.} Unless otherwise stated, the
superscript in the residual denotes the iteration number within the
Poisson solver.

\mbigskip

Then, recalling that $\diver \vel^{p}$ can be expressed as follows
\XaviR{(see, for instance, Batchelor~\cite{BAT51} or the textbook
  treatment in \bef{Chapter~2} \aft{Section~2.5} of
  Pope~\cite{POP00})}
\begin{equation}
\label{QGrel}
\diver \vel^{p} \approx \Dt \diver ( \vel \cdot \nabla \vel ) = 2 \Dt \QG ,
\end{equation}
\noindent leads to 
%
%
\begin{equation}
\label{residual3}
r^0 \approx 2 \Dt^{\scalResDt} \partial_t \QG \hphantom{kk} \text{with} \hphantom{kk}  \scalResDt=\left\{ \begin{array}{ll} 1 & \text{if $r^{0}$ defined as \Xavi{in} Eq.(\ref{residual1})} \\ 2 & \text{if $r^{0}$ defined as \Xavi{in} Eq.(\ref{residual2})} \end{array} \right.
\end{equation}
\noindent where $Q_G=-1/2 tr(\Gten^2)$ is the second invariant of the
velocity gradient tensor, $\Gten \equiv \nabla \vel$. Hence, smaller
$\Delta t$ reduces both $r^0$ and $\tilde{r}^0$, leading to faster
convergence. \aft{Notice that the exponent $\scalResDt$ simply
  reflects the normalization adopted for the Poisson equation with
  respect to the time-step, $\Dt$. For example, some formulations
  write $\lapl \press = \frac{1}{\Delta t} \diver \vel^{p}$
  ($\scalResDt=1$), while others absorb the time-step into the unknown
  by defining a pseudo-pressure, leading to $\lapl (\Dt\,\press) =
  \lapl \tilde{\press} = \diver \vel^{p}$ ($\scalResDt=2$). Since both
  normalizations are used in practice, we retain the exponent
  $\scalResDt$ in order to keep the formulation general and applicable
  to both normalizations.}

\mbigskip

At the same time, increasing $Re$ implies finer grids
(Eq.~\ref{K41_Nx}), \Xavi{resulting in a broader range of scales and,
  consequently, a more ill-conditioned Poisson equation.}
\XaviR{Therefore}, the spectral distribution of the residual
$\hat{r}_k^0$ is of central importance. \aft{Here and in the
  following, the hat notation $\hat{(\cdot)}$ denotes the Fourier
  coefficient of the corresponding quantity, while the sub-index $k$
  represents the associated wavenumber.} Assuming a power-law scaling
in the inertial range \XaviR{with slope $\scalResK$, we obtain},
\begin{equation}
\label{res_k}
\partial_t (\hat{Q}_\Gten)_k \propto k^{\scalResK}  \hspace{6.69mm} \Longrightarrow \hspace{6.69mm} \hat{r}_{k}^{0} \stackrel{(\ref{residual3})}{\approx} 2 \Dt^{\scalResDt} \partial_t (\hat{Q}_\Gten)_k \propto \Dt^{\scalResDt} k^{\scalResK} ,
\end{equation}
\noindent where $k$ is the wavenumber and $\scalResDt \in \{1,2\}$
depends on the definition of the residual: $\scalResDt=1$ for
Eq.~(\ref{residual1}) and $\scalResDt=2$ for
Eq.~(\ref{residual2}). \XaviR{This power-law scaling is confirmed
  \textit{a posteriori} by the numerical results presented in
  Section~\ref{results}.}

\mbigskip

A power-law scaling for $\QG$ can be derived from Eqs.(\ref{Poisson})
and~(\ref{QGrel}), and the $k^{-7/3}$ scaling of the shell-summed
squared pressure \XaviR{spectrum~\cite{BAT51,PUL95,CAO99,GOT99,GOT01,ZHA16,XU20}},
\begin{equation}
\label{QGscaling}
(\hat{Q}_\Gten)_k \propto k^{2} (k^{-7/3})^{1/2} = k^{5/6} .
\end{equation}
\aft{Notice that here the term pressure spectrum refers to the
  spectrum of the Fourier amplitudes, $|\press_k|$, whereas the
  squared magnitude, $|\press_k|^2$, is referred to explicitly as the
  squared pressure spectrum.}

\mbigskip

Then, the exponent value $\beta$ in Eq.(\ref{res_k}) can be inferred
from the dynamics of the invariants obtained from the so-called
restricted Euler equation~\cite{CAN92},
\begin{equation}
\label{REEq}
\partial_t \QG = - ( \vel \cdot \nabla ) \QG - 3 \RG ,
\end{equation}
\noindent where $\RG = det ( \Gten ) = 1/3 tr (\Gten^3)$ is the third
invariant of $\Gten$. \aft{Notice that the restricted Euler equation
  corresponds to a simplified description of the velocity-gradient
  dynamics in which viscous effects and the non-local pressure Hessian
  are neglected, retaining only the self-amplification associated with
  the nonlinear convective term. Then,} the two terms on the
right-hand side scale differently. Specifically,
\begin{align}
\label{convQscaling} (\widehat{( \vel \cdot \nabla ) \QG})_k &\propto (\widehat{ \nabla \QG })_k \propto k ( k^{5/6} ) = k^{11/6} , \\
\label{RGscaling} (\hat{R}_\Gten)_k &\propto ( k^{5/6} )^{3/2} = k^{5/4} 
\end{align}
\noindent \bef{where Taylor's frozen-turbulence
  hypothesis~\cite{TAY38} is applied to approximate the convective
  term, $( \vel \cdot \nabla ) \QG$, which is expected to become the
  dominant contribution on the right-hand side of Eq.(\ref{REEq})
  \Xavi{due to its steeper $k$-scaling}. Combining this with the
  results obtained in Eqs.(\ref{res_k}) and~(\ref{convQscaling}) leads
  to}

\noindent \aft{where Taylor's frozen-turbulence
  hypothesis~\cite{TAY38}, which assumes that turbulent structures are
  primarily advected by the large-scale velocity without significant
  distortion over short times, is invoked to estimate the spectral
  scaling of the convective term $(\vel \cdot \nabla)\QG$. Under this
  assumption, the action of the convective operator, $(\vel \cdot
  \nabla)$, on a transported quantity can be interpreted as a spatial
  derivative acting on that quantity, which introduces a factor
  proportional to the wavenumber~$k$ in Fourier space.}

\mbigskip

\aft{In view of these results, the convective term $(\vel \cdot
  \nabla)\QG$ is expected to become the dominant contribution on the
  right-hand side of Eq.~(\ref{REEq}) due to its steeper $k$-scaling.}
Combining this with the results obtained in Eqs.~(\ref{res_k})
and~(\ref{convQscaling}) leads to

\begin{equation}
\label{res_k2}
\hat{r}_{k}^{0} \propto \Dt^{\scalResDt} k^{\scalResK}  \hphantom{kkk} \text{with} \hphantom{kkk} \scalResK = 11/6 .
\end{equation}
Furthermore, we can assume that, given a flow configuration, the
proportionality constant, $\ResConst$, scales with the inverse of the
$Re$-number
\begin{equation}
\label{res_k3}
\hat{r}_{k}^{0} \approx \ResConst(Re) \Dt^{\scalResDt} k^{\scalResK} \propto Re^{-1} \Dt^{\scalResDt} k^{\scalResK}  \hphantom{kkk} \text{with} \hphantom{kkk} \scalResK = 11/6 .
\end{equation}
\noindent The reasoning behind this scaling is the following. Let us
consider a flow configuration with a forcing term that keeps the
energy of the largest scales constant, independently of the
$Re$-number. If we also assume that the flow is in equilibrium,
\ie~the energy distribution remains approximately constant over time,
then the non-linear convective term scales with the inverse of $Re$
\begin{equation}
\label{equilibrium}
\widehat{( \vel \cdot \nabla \vel)}_k \sim \frac{k^2}{Re} \hat{\vel}_k .
\end{equation}
\aft{Notice that this balance holds for the wavenumber range where the
  forcing term is absent, \ie~outside the range of large scales where
  the forcing is applied. Then,} plugging this into Eqs.(\ref{res_k})
and~(\ref{convQscaling}) and recalling the definition of the invariant
$\QG$, given in Eq.(\ref{QGrel}), leads to the conclusion that
$\ResConst(Re) \propto 1/Re$. Numerical tests with the 1D Burgers'
equation, presented in Section~\ref{Burgers}, confirm this scaling.

\mbigskip

Then, combining the results obtained in Eqs.(\ref{residual3})
and~(\ref{res_k3}) leads to
\begin{equation}
\label{res_k2}
\boxed{\hat{r}_{k}^{0} \propto Re^{-1} \Dt^{\scalResDt} k^{\scalResK}  \hphantom{kk} \text{with} \hphantom{kk} \scalResK = 11/6 \hphantom{kk} \text{and} \hphantom{kk} \scalResDt=\left\{ \begin{array}{lc} 1 & \text{if $\hat{r}$ defined as \Xavi{in} Eq.(\ref{residual1})} \\ 2 & \text{if $\hat{r}$ defined as \Xavi{in} Eq.(\ref{residual2})} \end{array} \right.}
\end{equation}
\XaviR{At this point, it is relevant to note that the derivation of
  this scaling relies on the classical $k^{-7/3}$ Kolmogorov scaling
  of the squared pressure spectrum in order to derive the scaling of
  the second invariant $Q_{\Gten}$ (see Eq.~\ref{QGscaling}). This
  scaling is, in principle, expected to hold only in the bulk region
  of turbulent flows, where assumptions of local homogeneity and
  isotropy are approximately satisfied. In wall-bounded
  configurations, alternative scalings are predicted for the
  logarithmic layer, most notably a $k^{-1}$ behavior arising from
  attached-eddy arguments~\cite{TOW76}, that has been reported in both
  experimental measurements~\cite{TSU07} and numerical
  simulations~\cite{PAT14,XU20,PIR25} of near-wall pressure
  fluctuations. Such deviations from the pressure scaling in the bulk
  region suggest that the dynamics of the invariant $Q_{\Gten}$ may
  differ in the near-wall region, as supported by studies reporting
  relevant changes in the invariant-based analysis of the flow
  topology in the near-wall region~\cite{BEC17,YIG20}. This, in turn,
  may potentially lead to a modified value of the exponent $\scalResK$
  and, consequently, to different scaling trends for the Poisson
  solver residual. Nevertheless, the numerical results for
  wall-bounded turbulent flows presented in
  Section~\ref{complex_flows} support the idea that the effective
  value of $\scalResK$ remains practically unchanged when compared to
  bulk turbulence. This behavior may be attributed to the inherently
  non-local nature of the pressure Poisson equation, which involves
  long-range interactions between outer-layer motions and the
  near-wall region that are not directly damped by
  viscosity~\cite{PIR25}.}

\mbigskip

\noindent In summary, there are two competing effects (see
Figure~\ref{Competing_effects}) when increasing the $Re$ number: the
time-step, $\Dt$, and the proportionality constant decrease whereas
the range of scales \Xavi{increases}. The next step is to analyze how
the solver convergence is affected.

\begin{figure}[!t]
  \begin{center}
    \includegraphics[width=0.57\textwidth]{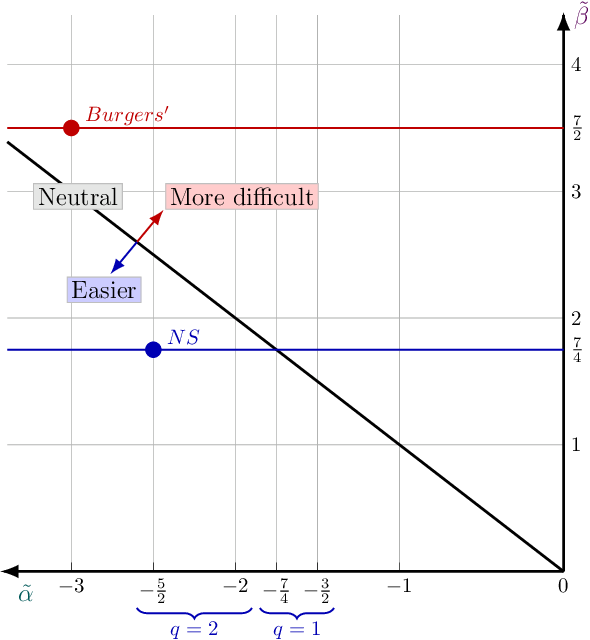}
  \end{center}
\vspace{-3mm}
\caption{Phase space $\{ \scalResReyNum , \scalResReyPhy \}$. Solid
  black line corresponds to $|| r^n ||^2 \propto Re^{0}$ in
  Eqs.(\ref{Jacobi_residual_scaling}) and~(\ref{MG_residual_scaling}),
  \ie~neutral effect of $Re$-number in the total number of iterations,
  and corresponds to $\scalResReyNum = - \scalResReyPhy$. Horizontal
  blue line corresponds to $\scalResReyPhy=7/4$ which is the
  estimation for the NS equations. The blue dot labeled as NS
  corresponds to the most common situation where $q=2$ (see
  Eq.~\ref{residual3}) and $\scalDtRey=-3/4$ (see
  Eq.~\ref{Dt_scaling}) leading to $\scalResReyNum=-5/2$ (see
  Eq.~\ref{alphatilde_def}). The horizontal red line corresponds to
  the same analysis but for the Burgers' equation studied in
  Section~\ref{results}.}
\label{Phase_space}
\end{figure}

\section{ANALYSIS OF THE SOLVER CONVERGENCE}

\label{solver}

We want to study whether the number of iterations inside the Poisson's
solver increases or decreases with $Re$. To do so, we can relate the
L2-norm of the residual with the integral of $\hat{r}_{k}^{2}$ for all
the wavenumbers using the Parseval's theorem, \ie
\begin{equation}
\label{Parseval}
|| r ||^2 = \int_{\Omega} r^2 \ud V = \int_{1}^{k_{\max}} \hat{r}_{k}^{2} \ud k ,
\end{equation}
\noindent where $k_{\max} \sim 1 / \eta$ is the maximum wavenumber and
$\eta$ is the smallest resolved scale. Therefore, a power-law relation
exists between $k_{\max}$ and the Reynolds number,
\begin{equation}
\label{kmax_scal}
k_{\max} \approx C_{k} Re^{\scalKmaxRey} ,
\end{equation}
\noindent where, for the \Xavi{NS~equations}, the exponent is
$\scalKmaxRey=3/4$ (see Eq.~\ref{K41_Nx}). \XaviR{Nevertheless, we
  retain the general form given in Eq.~(\ref{kmax_scal}), since the
  Burgers’ equation, characterized by a different value of
  $\scalKmaxRey$ (see Table~\ref{scalings}), is also examined
  numerically in the next section.}

\mbigskip

Then, the residual after $n$ iterations can be computed as
\begin{equation}
|| r^n ||^2 = \int_{1}^{k_{\max}} \left( \hat{\omega}_{k}^{n} \hat{r}_{k}^{0} \right)^2 \ud k \stackrel{(\ref{Dt_scaling}) (\ref{res_k2})\XaviR{(\ref{kmax_scal})}}{\approx} \int_{1}^{\KmaxConst Re^{\scalKmaxRey}} \hat{\omega}_{k}^{2n} Re^{2 \tilde{\alpha}} k^{2 \beta} \ud k ,
\end{equation}
\noindent where \aft{$\hat{\omega}_{k} =
  \hat{r}_{k}^{n+1}/\hat{r}_{k}^{n}$ is the convergence ratio of the
  solver and}
\begin{equation}
\label{alphatilde_def}
\tilde{\alpha} \equiv \scalResDt \alpha -1 ,
\end{equation}
\noindent \bef{and $\hat{\omega}_{k} =
  \hat{r}_{k}^{n+1}/\hat{r}_{k}^{n}$ is the convergence ratio of the
  solver} \aft{where $\alpha$ and $\scalResDt$ are defined in
  Eqs.(\ref{Dt_scaling}) and~(\ref{residual3}), respectively}. For
instance, for a Jacobi solver, $\hat{\omega}_{k} = \cos (
\frac{\pi}{2} \rho)$ where $\rho \equiv k / k_{\max}$, \Xavi{which
  corresponds to the classical second-order finite-difference (also
  finite-volume) discretization of the Poisson equation on a
  \XaviR{uniform grid~\cite{BRI87}}; other discretizations lead to
  different expressions but exhibit the same qualitative behavior.} In
this case, using a quadratic approximation of $\cos (x) \approx 1 - 4
x^2 / \pi^2$ and applying the change of variable $k=k_{\max} \rho
\approx \KmaxConst Re^{\scalKmaxRey} \rho$ \XaviR{(see
  Eq.~\ref{kmax_scal})} leads to
\begin{equation}
|| r^n ||^2 \approx \KmaxConst^{2\scalResK+1} Re^{2 (\scalResReyNum + \scalResReyPhy )} \int_{1/k_{\max}}^{1} \rho^{2\scalResK}  (1 - \rho^2)^{2n} \ud \rho ,
\end{equation}
\noindent where 
\begin{equation}
\label{betatilde_def}
\scalResReyPhy \equiv \scalKmaxRey \left( \scalResK + \frac{1}{2} \right) .
\end{equation}
\noindent Therefore, $\scalResReyNum$ in Eq.(\ref{alphatilde_def})
represents the part of the residual \XaviR{that scales} with $Re$ that
is associated with numerical aspects, whereas $\scalResReyPhy$ is
determined by the underlying flow physics. For instance, in the case
of the NS equations, $\scalKmaxRey=3/4$ (see Eq.\ref{K41_Nx}) and
$\scalResK=11/6$ (see Eq.~\ref{res_k2}), leading to $\scalResReyPhy =
7/4$.

\mbigskip

\noindent Then, assuming that $k_{\max} \gg 1$, the integral can be
accurately approximated by 
\begin{equation}
\label{Jacobi_residual_betaf}
|| r^n ||^2 \approx \KmaxConst^{2\scalResK+1} Re^{2 (\scalResReyNum + \scalResReyPhy)} \frac{1}{2} \mathcal{B} ( \scalResK + 1/2 , 2n+1 ) .
\end{equation}

\noindent where $\mathcal{B}(a,b) = \int_{0}^{1} t^{a-1} (1-t)^{b-1}
\ud t$ is the beta function. Finally, assuming that $n \gg 1$, the
beta function can be approximated as follows
\begin{equation}
|| r^n ||^2 \approx \KmaxConst^{2\scalResK+1} Re^{2 (\scalResReyNum + \scalResReyPhy)} \frac{1}{2} \Gamma(\scalResK + 1/2 ) (2n+1)^{-(\scalResK+1/2)} ,
\end{equation}
\noindent whereas for the initial residual, $n=0$, it \XaviR{can be
  approximated as}
\begin{equation}
|| r^0 ||^2 \approx \KmaxConst^{2\scalResK+1} Re^{2 (\scalResReyNum + \scalResReyPhy)} \frac{1}{2} \frac{\Gamma(\scalResK + 1/2 )}{\Gamma(\scalResK + 3/2 )} ,
\end{equation}
\noindent where $\Gamma(\cdot)$ is the gamma function. In summary,
the L2-norm of the residual scales with $Re$ and convergences
exponentially
\begin{equation}
\label{Jacobi_residual_scaling}
|| r^n ||^2 \propto \frac{Re^{2 (\scalResReyNum + \scalResReyPhy)}}{(2n+1)^{\scalResK+1/2}} .
\end{equation}

Furthermore, from the expression given in
Eqs.(\ref{Jacobi_residual_scaling}), we can deduce how the total
number of solver iterations, $n$, scales with the
$Re$-number. Namely, $n \propto Re^{\scalnRey}$
\begin{equation}
\label{xi_def}  
Re^{2 (\scalResReyNum + \scalResReyPhy)} \sim (2n+1)^{\scalResK+1/2} \propto n^{\scalnRey (\scalResK+1/2)} \hphantom{kk} \Longrightarrow \hphantom{kk} \scalnRey = \frac{2 (\scalResReyNum + \scalResReyPhy)}{\scalResK + 1/2} .
\end{equation}
\noindent Theoretical results for this $Re$-scaling are shown in the
last column of Table~\ref{scalings} for both the NS and the Burgers'
equation. \XaviR{The latter case will be analyzed in detail in
  Section~\ref{Burgers}.}


%


\subsection{Extension to multigrid}

\label{MG_solver}

\noindent We can extend this analysis for \XaviR{an MG solver} with
$l_{\max} \sim \log_{2} N_x \sim \scalKmaxRey \log_{2} Re$ levels
\XaviR{(see Eq.~\ref{kmax_scal})} and the Jacobi as smoother at each
level. We also assume that the mesh resolution becomes twice coarser
at each level. Then, the $L2$-norm of the initial residual is
distributed as follows
\begin{align}
\nonumber || r^0 ||^2 &= \int_{k_{\max}/2}^{k_{\max}} (\hat{r}_{k}^{0})^2 \ud k + \int_{k_{\max}/4}^{k_{\max}/2} (\hat{r}_{k}^{0})^2 \ud k + \cdots + \int_{k_{\max}/2^{l_{\max}+1}}^{k_{\max}/2^{l_{\max}}} (\hat{r}_{k}^{0})^2 \ud k + \int_{1}^{k_{\max}/2^{l_{\max}+1}} (\hat{r}_{k}^{0})^2 \ud k \\
\label{res_split_MG} &= R_{0} = R_{0,1} + R_{1,2} + + \cdots + R_{l_{\max}-1,l_{\max}} + R_{l_{\max}} ,
\end{align}
\noindent where
\begin{equation}
R_{l_1,l_2} \equiv \int_{k_{\max}/2^{l_{2}}}^{k_{\max}/2^{l_{1}}} (\hat{r}_{k}^{0})^2 \XaviR{\ud k} \hphantom{kkk} \text{and} \hphantom{kkk} R_{l} \equiv \int_{1}^{k_{\max}/2^{l}} (\hat{r}_{k}^{0})^2 \XaviR{\ud k},
\end{equation}
\noindent represent the part of the $L2$-norm of the initial residual
contained between the wavenumbers $k_{\max}/2^{l_{2}}$ and
$k_{\max}/2^{l_{1}}$, and below the wavenumber $k_{\max}/2^{l}$,
respectively. Moreover, for the initial level, $l=0$, we can
accurately approximate the effective \XaviR{damping, given in
  Eq.(\ref{Jacobi_residual_betaf}),} of the residual as follows
\begin{equation}
|| r^n ||^2 \approx \KmaxConst^{2\scalResK+1} Re^{2 (\scalResReyNum + \scalResReyPhy)} \frac{1}{2} ( 1 - \mathcal{B}_{1/2} ( \scalResK + 1/2 , 2n+1 ) ), 
\end{equation}
\noindent where $\mathcal{B}_{x} (a,b) = \int_{0}^{x} t^{a-1}
(1-t)^{b-1}$ is the incomplete beta function. Recalling that
$\incbetaf{\half}{a}{b} = ( 1 - \incbetaf{\half}{b}{a})$, we can write
the previous expression more compactly as
\begin{equation}
|| r^n ||^2 \approx \KmaxConst^{2\scalResK+1} Re^{2 (\scalResReyNum + \scalResReyPhy)} \frac{1}{2} ( \mathcal{B}_{1/2} ( 2n+1 , \scalResK + 1/2 ) ) .
\end{equation}
The underlying idea is that the Jacobi smoother effectively damps only
the high-frequency components of the error. Consequently, its
influence is evaluated over the range $k_{\max}/2 < k \le k_{\max}$,
while its impact at lower wavenumbers is deliberately
neglected. Therefore, it is only damping $R_{0,1}$. The same logic
applies to all the subsequent MG levels up to $l_{\max}-1$ leading to
\begin{align}
  || r^n ||^2 &\approx \left( \sum_{l=0}^{l_{\max}-1} R_{l,l+1} \right) \frac{1}{2} \incbetaf{\half}{2n+1}{\scalResK + \half} + \frac{R_{l_{\max}}}{2} \betaf{2n+1}{\scalResK + \half} \\
&\stackrel{(\ref{res_split_MG})}{=} \frac{R_0 - R_{l_{\max}}}{2} \incbetaf{\half}{2n+1}{\scalResK + \half} + \frac{R_{l_{\max}}}{2} \betaf{2n+1}{\scalResK + \half} \XaviR{.}
\end{align}
\noindent Notice that the last level, $l_{\max}$, is also solved using
a Jacobi solver. In a practical MG implementation, this last level is
usually solved with a direct solver or, at least, with a more
efficient solver~\cite{VacCarSoo24}.

\mbigskip

Recalling the definition of the \XaviR{residual, given in
  Eq.(\ref{Jacobi_residual_betaf}),} it leads to
\begin{equation}
\label{MG_residual_scaling}
|| r^n ||^2 \approx \KmaxConst^{2\scalResK+1} Re^{2 (\scalResReyNum + \scalResReyPhy)} \left\{\left( 1 - \frac{R_{l_{\max}}}{R_0} \right) \frac{\incbetaf{\half}{2n+1}{\scalResK + \half}}{2} + \frac{R_{l_{\max}}}{R_0} \frac{\betaf{2n+1}{\scalResK + \half}}{2} \right\} .
\end{equation}
\noindent Compared to Eq.(\ref{Jacobi_residual_betaf}), MG is strongly
accelerated by the term in brackets. Moreover, notice that if
$l_{\max}=0$, \ie~zero MG level, it collapses to the formula derived
for the \XaviR{Jacobi-only} solver. Nevertheless, the scaling with
$Re$ is the same; therefore, the regions defined in the $\{
\scalResReyNum, \scalResReyPhy \}$ phase space remain unchanged (see
Figure~\ref{Phase_space}).

\mbigskip

\newcommand{\myfrac}[2]{{#1}/{#2}}

\begin{table}[!t]
\begin{center}
\begin{tabular}{l|ccc|ccc|c}
                          & $\scalDtRey$                  & $\scalResDt$         & $\scalResReyNum$          & $\scalResK$       & $\scalKmaxRey$       & $\scalResReyPhy$         & $\scalnRey$ \\
\hline
\hline
Formula                   & Eq.(\ref{Dt_scaling})         & Eq.(\ref{residual3}) & Eq.(\ref{alphatilde_def}) & Eq.(\ref{res_k2})         & Eq.(\ref{kmax_scal}) & Eq.(\ref{betatilde_def}) & Eq.(\ref{xi_def}) \\
\hline 
\multirow{2}{*}{NS}       & $-\myfrac{3}{4}$              & $2$                  & $-\myfrac{5}{2}$          & $\myfrac{11}{6}$          & $\myfrac{3}{4}$      & $\myfrac{7}{4}$          & $-\myfrac{9}{14}$ \\
                          & $-\myfrac{3}{4}$              & $1$                  & $-\myfrac{7}{4}$          & $\myfrac{11}{6}$          & $\myfrac{3}{4}$      & $\myfrac{7}{4}$          & $0$ \\
\hline
\hline
Formula                   & Eq.(\ref{Dt_scaling_Burgers}) & Eq.(\ref{residual3}) & Eq.(\ref{alphatilde_def}) & Eq.(\ref{res_k2_Burgers}) & Eq.(\ref{kmax_scal}) & Eq.(\ref{betatilde_def}) & Eq.(\ref{xi_def}) \\
\hline
\multirow{2}{*}{Burgers'} & $-1$                          & $2$                  & $-3$                      & $3$                       & $1$                  & $\myfrac{7}{2}$          & $\myfrac{2}{7}$ \\
                          & $-1$                          & $1$                  & $-2$                      & $3$                       & $1$                  & $\myfrac{7}{2}$          & $\myfrac{6}{7}$
\end{tabular}
\end{center}
\caption{Exponents for all the relevant scalings for both the NS and
  the Burgers' equation.}
\label{scalings}
\end{table}

\newcommand{\spectrawidth}{0.49\textwidth}
\newcommand{\largespectrawidth}{0.69\textwidth}
\newcommand{\DNSflowswidth}{0.69\textwidth}


\begin{figure}[!t]
\begin{center}
\includegraphics[angle=-90,width=\spectrawidth]{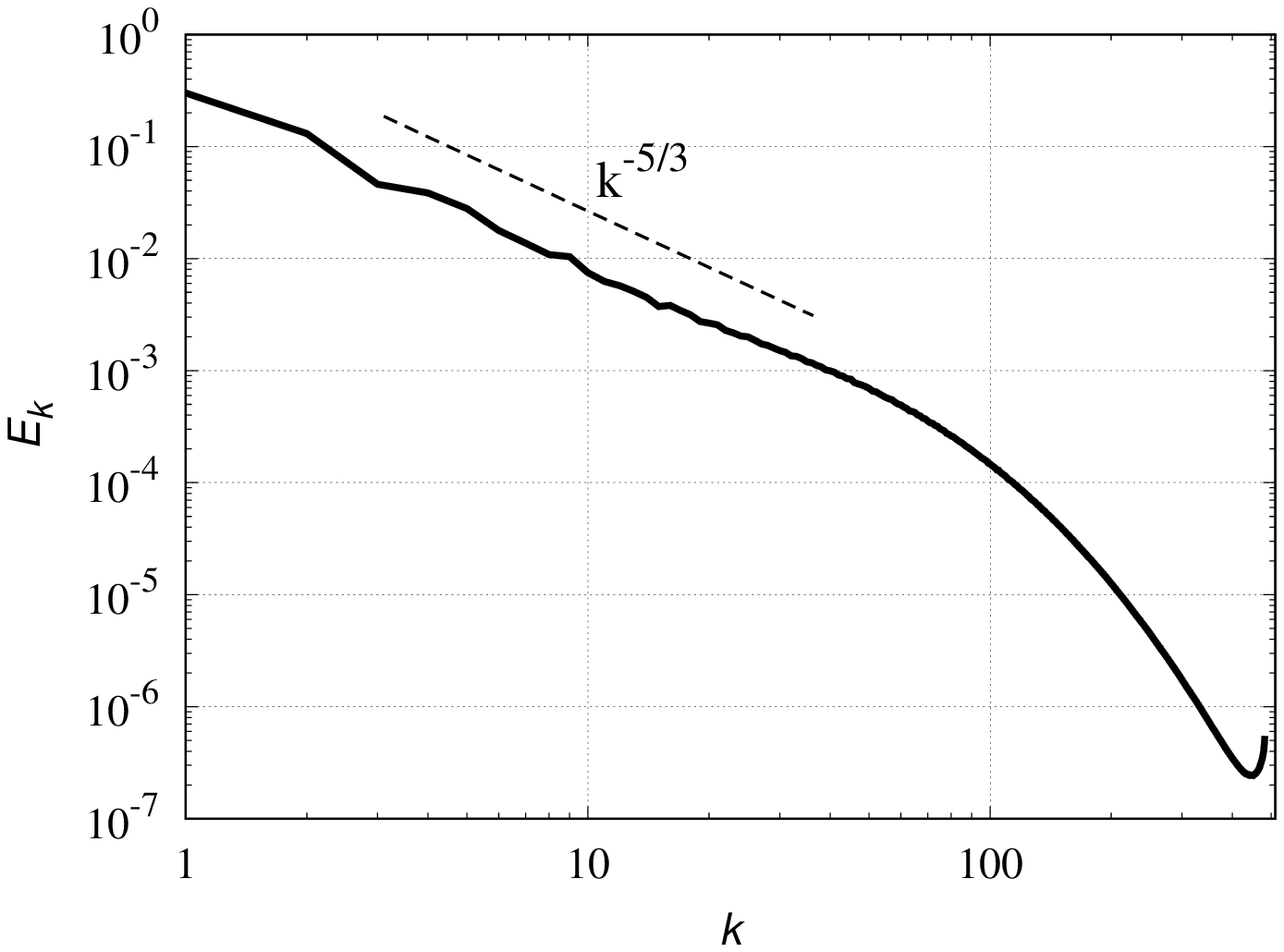}
\includegraphics[angle=-90,width=\spectrawidth]{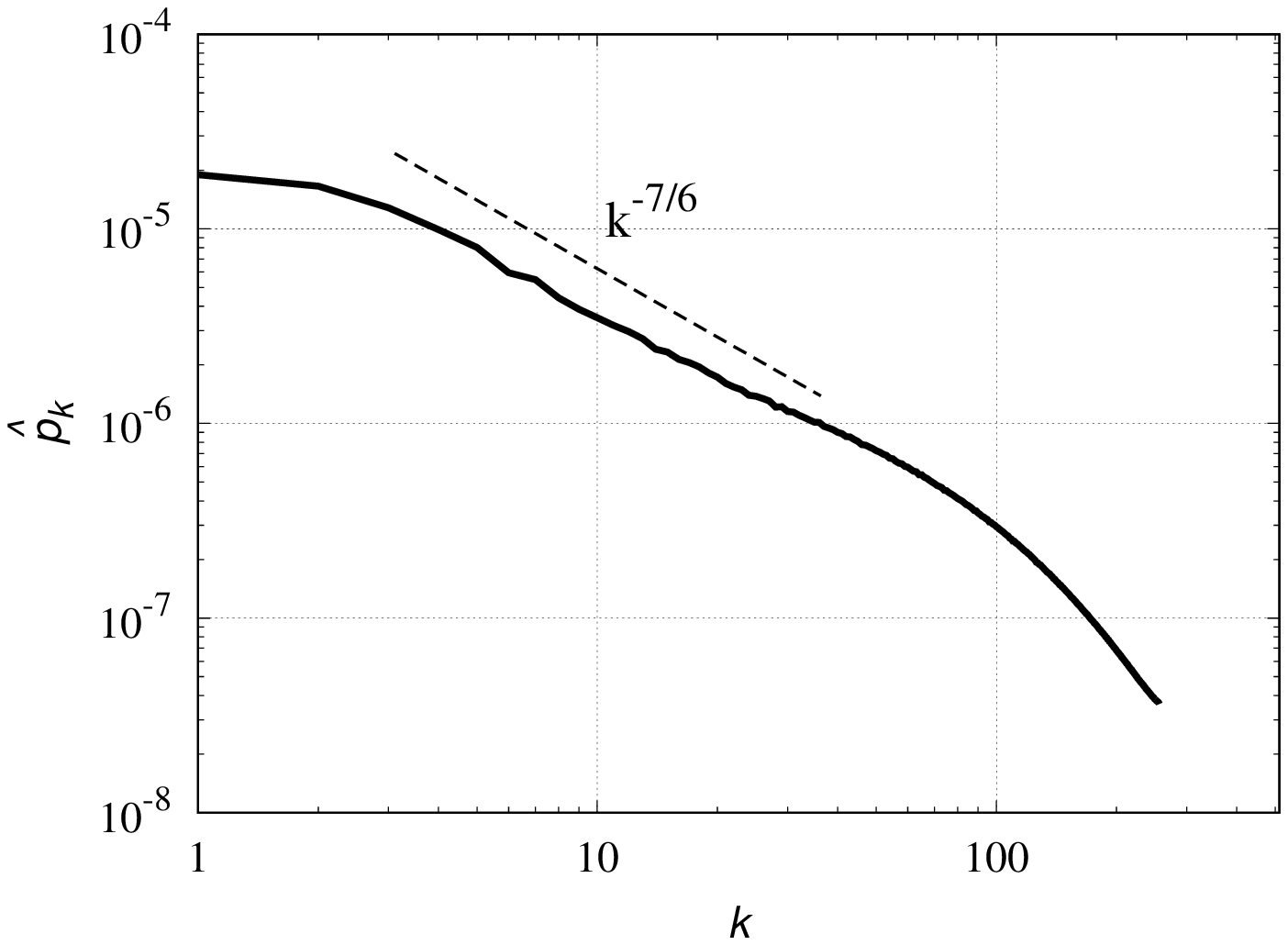}
\end{center}
\caption{Energy and pressure spectra for the forced HIT simulation at
  $Re_\lambda \approx 433$. Data has been obtained from the JHTDB
  database~\cite{PER07-JHTDB,LI08-JHTDB}.}
\label{Energy_and_pressure_spectra}
\end{figure}

\begin{figure}[!t]
\begin{center}
\includegraphics[angle=-90,width=\spectrawidth]{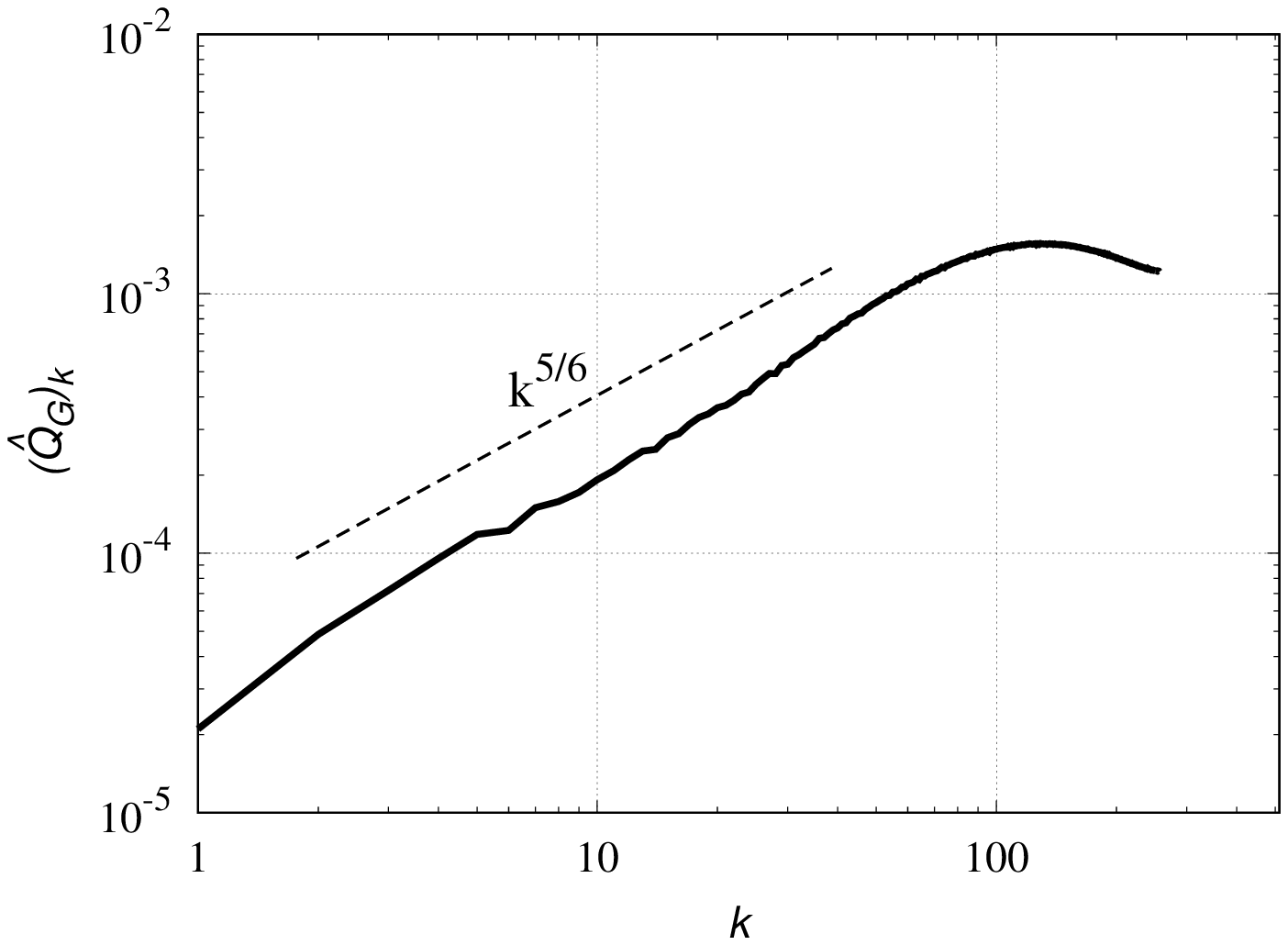}
\includegraphics[angle=-90,width=\spectrawidth]{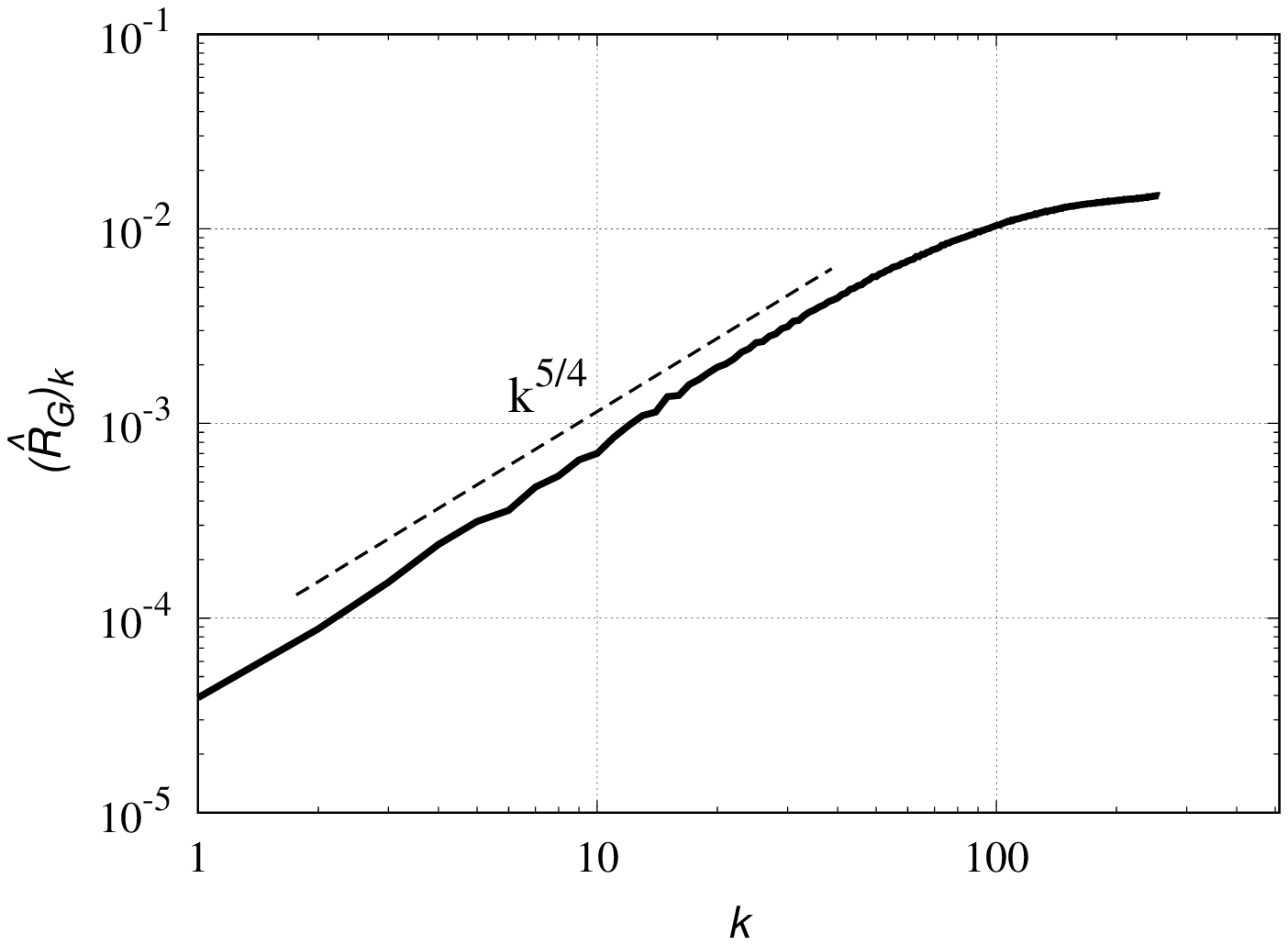}
\end{center}
\caption{Same as in Figure~\ref{Energy_and_pressure_spectra} but
for the second, $\QG$, and third invariant, $\RG$, of the velocity
gradient tensor.}
\label{QR_spectra}
\end{figure}

\begin{figure}[!t]
\begin{center}
\includegraphics[angle=-90,width=\spectrawidth]{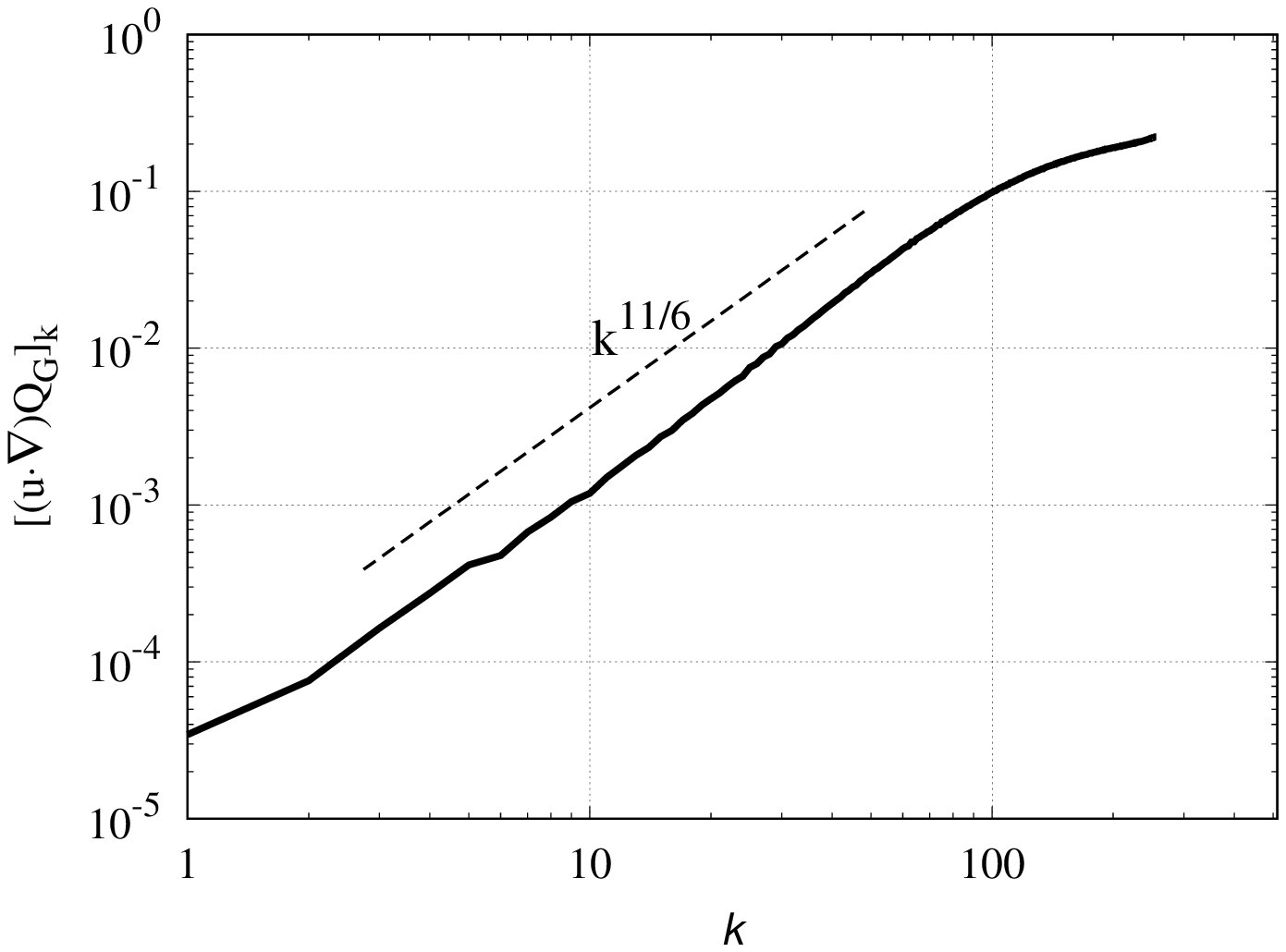}
\includegraphics[angle=-90,width=\spectrawidth]{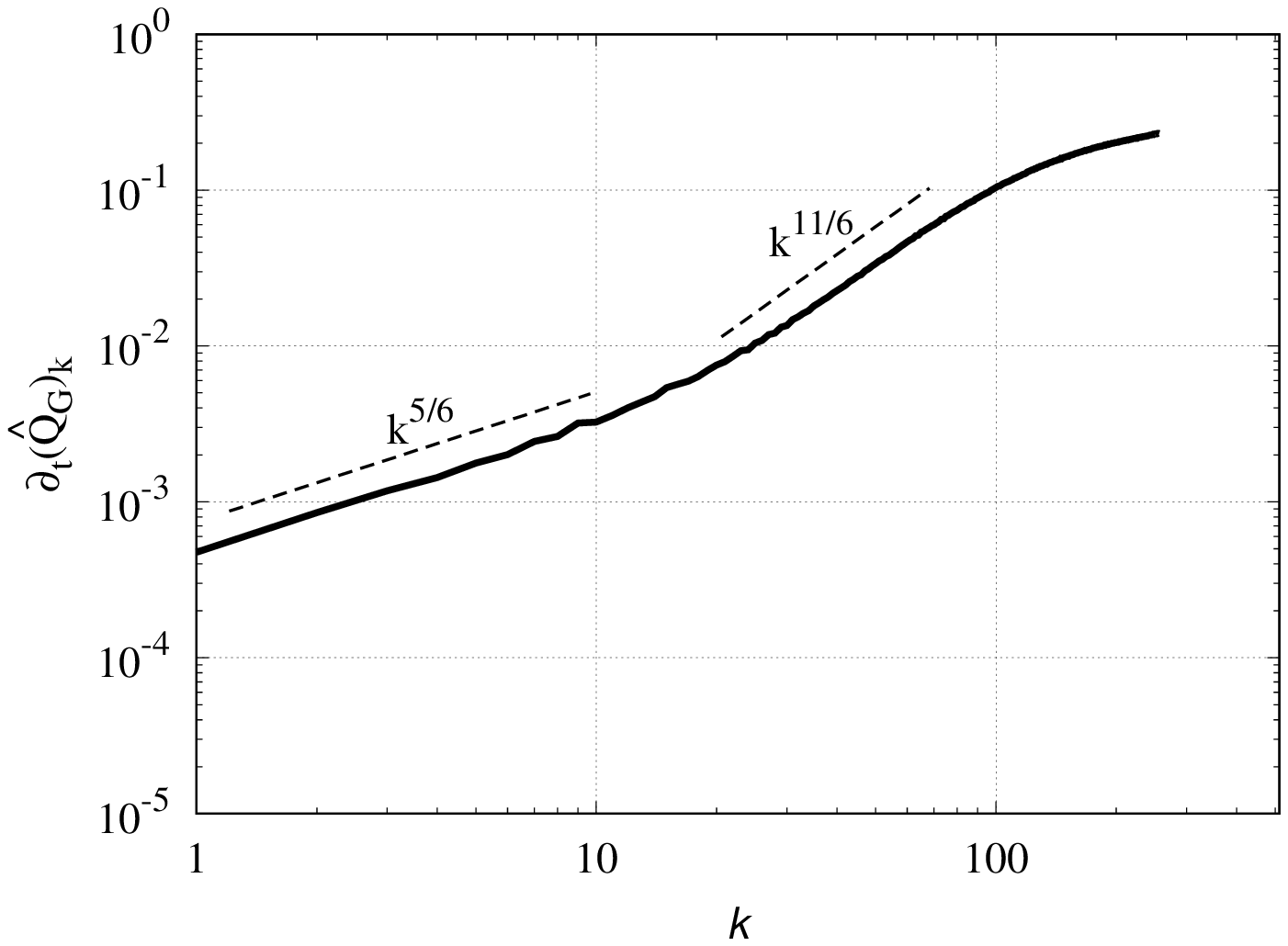}
\end{center}
\caption{Same as in Figure~\ref{Energy_and_pressure_spectra} but
for the convective term, $( \vel \cdot \nabla) \QG$, and the residual
of the Poisson equation.}
\label{Conv_and_residual_spectra}
\end{figure}

\begin{figure}[!t]
\begin{center}
\includegraphics[angle=-90,width=\largespectrawidth]{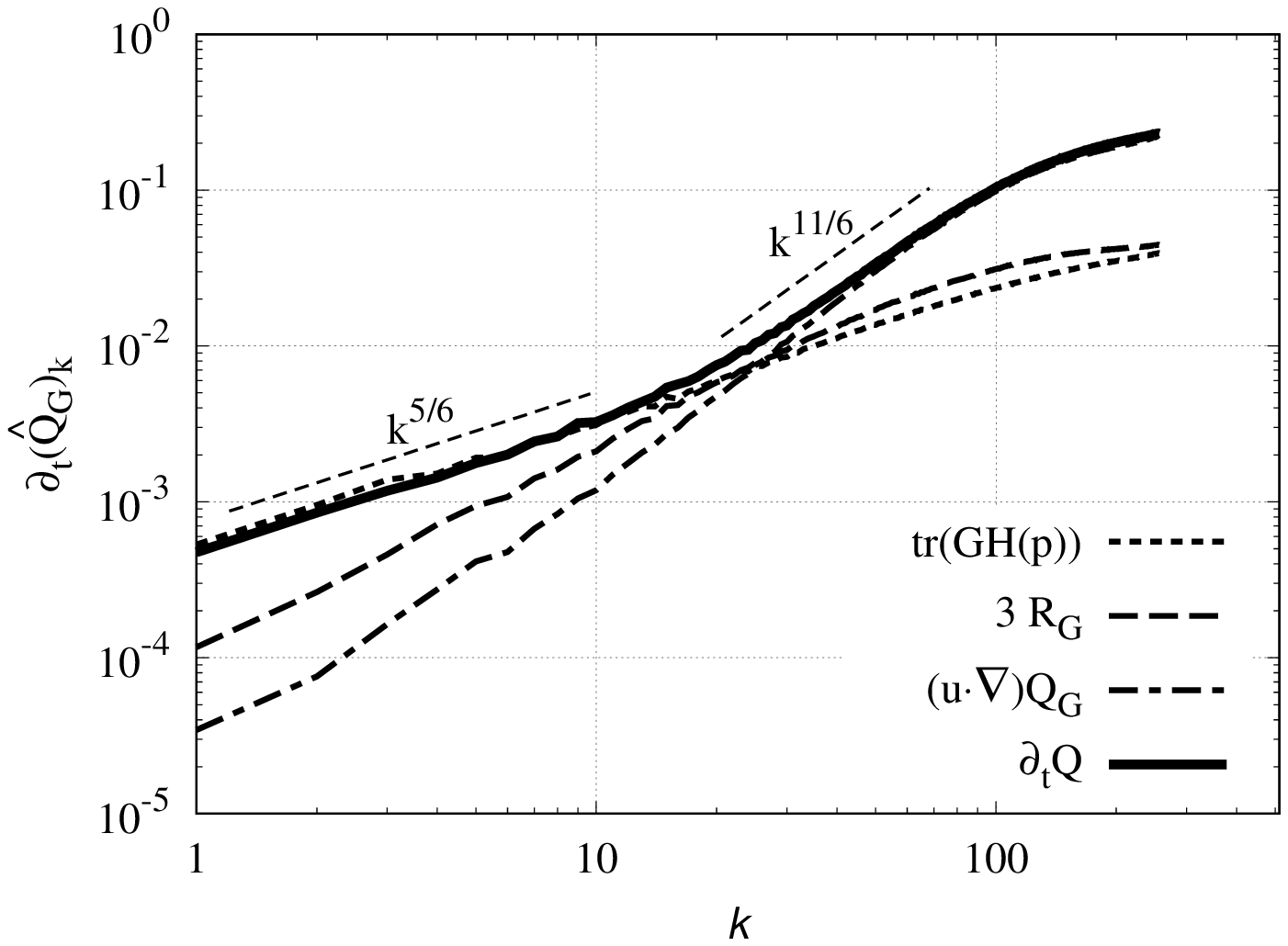}
\end{center}
\caption{Same as in Figure~\ref{Energy_and_pressure_spectra} but for $\partial_t \QG$
and its breakdown into the different terms that contribute to it in
Eq.(\ref{DtQG}).}
\label{Residual_breakdown_spectrum}
\end{figure}


\section{Numerical results}

\label{results}

\subsection{Homogeneous isotropic turbulence}

\label{HIT}

As a first validation case, we consider the forced HIT using
\XaviR{the} data from the Johns Hopkins Turbulence Database
(JHTDB)~\cite{PER07-JHTDB,LI08-JHTDB}. The chosen dataset corresponds
to a DNS of forced isotropic turbulence at Taylor microscale Reynolds
number $Re_\lambda \approx 433$, carried out on a $1024^3$ grid. The
flow is statistically stationary due to a large-scale forcing that
maintains a constant energy level in shells such that $k \le 2$. The
database provides access to velocity and pressure fields as well as
all spatial derivatives over a sequence of consecutive time steps,
which enables the computation of spectra of various quantities of
interest, including the principal invariants of the velocity gradient
tensor and their temporal derivatives. Further details regarding the
numerical setup and simulation methodology can be found in the
original publication\cite{LI08-HIT}.

\mbigskip

The homogeneous and isotropic nature of the flow makes this dataset an
ideal benchmark for assessing the spectral scaling laws derived in
Section~\ref{residual}. As expected from \XaviR{the} classical
turbulence theory, the kinetic energy spectrum reproduces the
well-known Kolmogorov $k^{-5/3}$~scaling, while the pressure spectrum
exhibits \XaviR{the
  $k^{-7/6}$~slope~\cite{BAT51,PUL95,CAO99,GOT99,GOT01,ZHA16,XU20}.}
These results, which are shown in
Figure~\ref{Energy_and_pressure_spectra}, are fully consistent with
previous studies and confirm that the database accurately reproduces
the universal features of isotropic turbulence across the inertial
range. Then, \Xavi{the} spectra for the second, $\QG$, and third,
$\RG$, principals invariants of the velocity gradient tensor are
displayed in Figure~\ref{QR_spectra}. The power-law scalings are also
consistent with the predictions given in Eqs.(\ref{QGscaling})
and~(\ref{RGscaling}), respectively. This confirms the assumptions
made in Section~\ref{Re_scalings} that eventually lead to the
$\scalResK = 11/6$ scaling of the residual (see
Eq.~\ref{res_k2}). \Xavi{This scaling follows from that of the
  convective term, $(\vel \cdot \nabla) \QG$, given in the
  Eq.(\ref{REEq}), which is derived from the restricted Euler
  equation.} The $k^{11/6}$ scaling of this term is confirmed in
Figure~\ref{Conv_and_residual_spectra} (left). Nevertheless, the
spectrum of the $\partial_t \QG$ shown in
Figure~\ref{Conv_and_residual_spectra}~(right) shows two regions:
namely, the predicted $11/6$~scaling at high wavenumbers, \Xavi{and a
  $5/6$~scaling at lower wavenumbers}. This second scaling cannot be
explained with the simplified model given in Eq.(\ref{REEq}). Namely,
the invariant $\RG$, which is the second term in the \Xavi{right-hand
  side} of the equation, scales with $k^{5/4}$ as shown in
Figure~\ref{QR_spectra} (right) and may have relevance only at low
\Xavi{wavenumbers}. However, it \Xavi{does not} explain the
$5/6$~scaling observed in
Figure~\ref{Conv_and_residual_spectra}~(right). To do so, we need a
more complete model.

\mbigskip

From the NS equations~(\ref{NS_eqs}), we can derive all the terms that
contribute to the evolution of the invariant $\QG$,
\begin{equation}
\label{DtQG}
\partial_t \QG = - ( \vel \cdot \nabla ) \QG - 3 \RG + tr(\Gten H_p) - \frac{1}{Re} tr (\Gten \lapl \Gten) ,
\end{equation}
\noindent where $H_p \equiv \nabla \nabla \press$ is the Hessian of
\Xavi{the} pressure \Xavi{field}. \aft{This equation makes explicit
  the viscous and pressure contributions that were neglected in the
  restricted Euler framework discussed above (see Eq.~\ref{REEq}). It
  therefore provides a useful reference to interpret the scaling
  arguments derived previously when the full dynamics of the system
  are considered.} The last term \aft{in Eq.(\ref{DtQG})} represents
the viscous effects, which are expected to have a relevant
contribution only in the dissipation range but not in the inertial
one. Therefore, we can restrict our analysis to the first three terms
in the \Xavi{right-hand side} of Eq.(\ref{DtQG}). Results are
displayed in Figure~\ref{Residual_breakdown_spectrum}. Firstly, we can
confirm the dominance of the convective term $(\vel \cdot \nabla )
\QG$ at high wavenumbers leading to the anticipated
$11/6$~scaling. Secondly, we can now explain the $5/6$~scaling
observed at low wavenumbers which is essencially due to the pressure
effects through the term $tr(\Gten H_p)$ in Eq.(\ref{DtQG}).
 
\mbigskip

In summary, the residual of the Poisson equation, $r^0$, which is
proportial to $\partial_t \QG$ as shown in Eq.(\ref{residual3}), has
two relevant contributions: $(\vel \cdot \nabla) \QG$ and $tr(\Gten
H_p)$, which correspond to the first and third term in the
right-hand-side of Eq.(\ref{DtQG}), respectively. The latter scales
with $k^{5/6}$ as shown in Figure~\ref{Residual_breakdown_spectrum}
and is relevant only at low wavenumber, whereas the former scales with
$k^{11/6}$ and eventually becomes the dominant at higher wavenumbers
confirming the adequacy of the analysis done in
Section~\ref{Re_scalings}. Nevertheless, two crucial issues remain to
be demonstrated: (i) whether the proposed theory also applies to
complex, non-homogeneous turbulent flows beyond HIT, and (ii) whether
the predicted scalings of both the solver residual and the iteration
count required to solve the Poisson equation are confirmed
numerically. These two points are addressed in the next subsections.

\begin{figure}[!t]
\begin{center}
    \includegraphics[width=\DNSflowswidth]{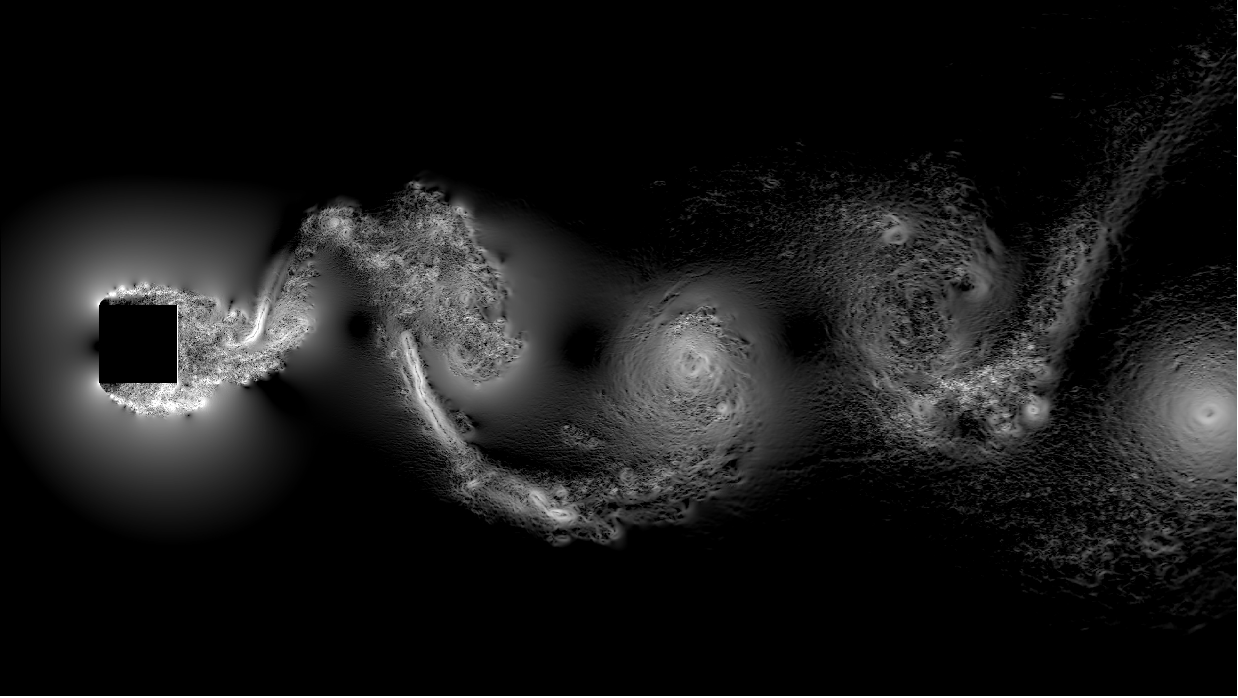}
\end{center}
\vspace{-1.69mm}
\begin{center}
    \includegraphics[width=\DNSflowswidth]{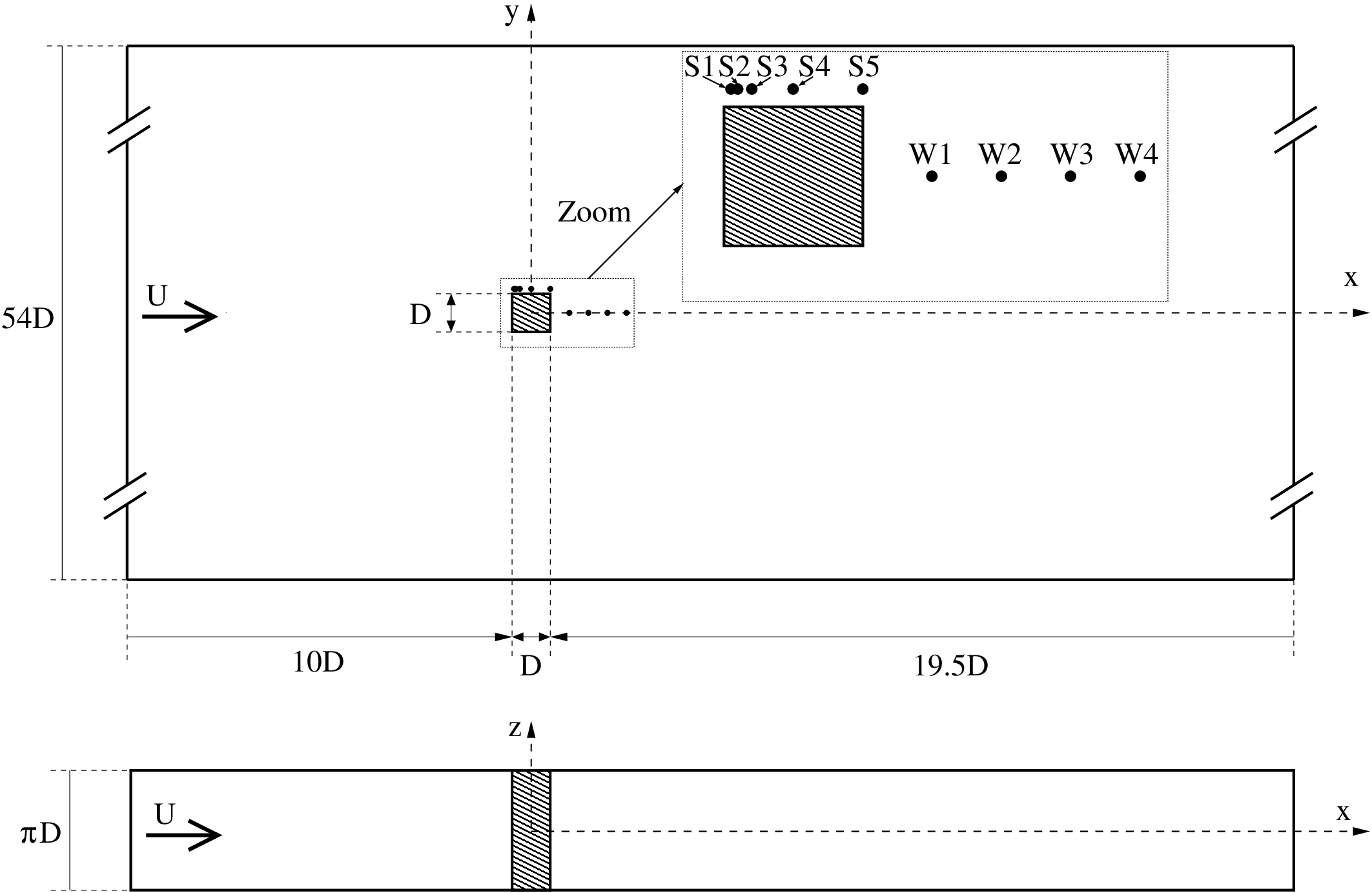}
\end{center}
\caption{\aft{Illustrative snapshot of the DNS simulation of turbulent
    flow around a square cylinder at $Re=55000$, together with a
    schematic representation of the computational domain and the
    location of the monitoring probes used in the analysis (see
    Table~\ref{points_CYL}). The simulation was performed on the
    MareNostrum~5-GPP supercomputer using $3136$~CPU cores with a mesh
    of $2.6$~billion grid points.}}
\label{CYL}
\end{figure}

\begin{figure}[!t]
\begin{center}
    \includegraphics[width=\DNSflowswidth]{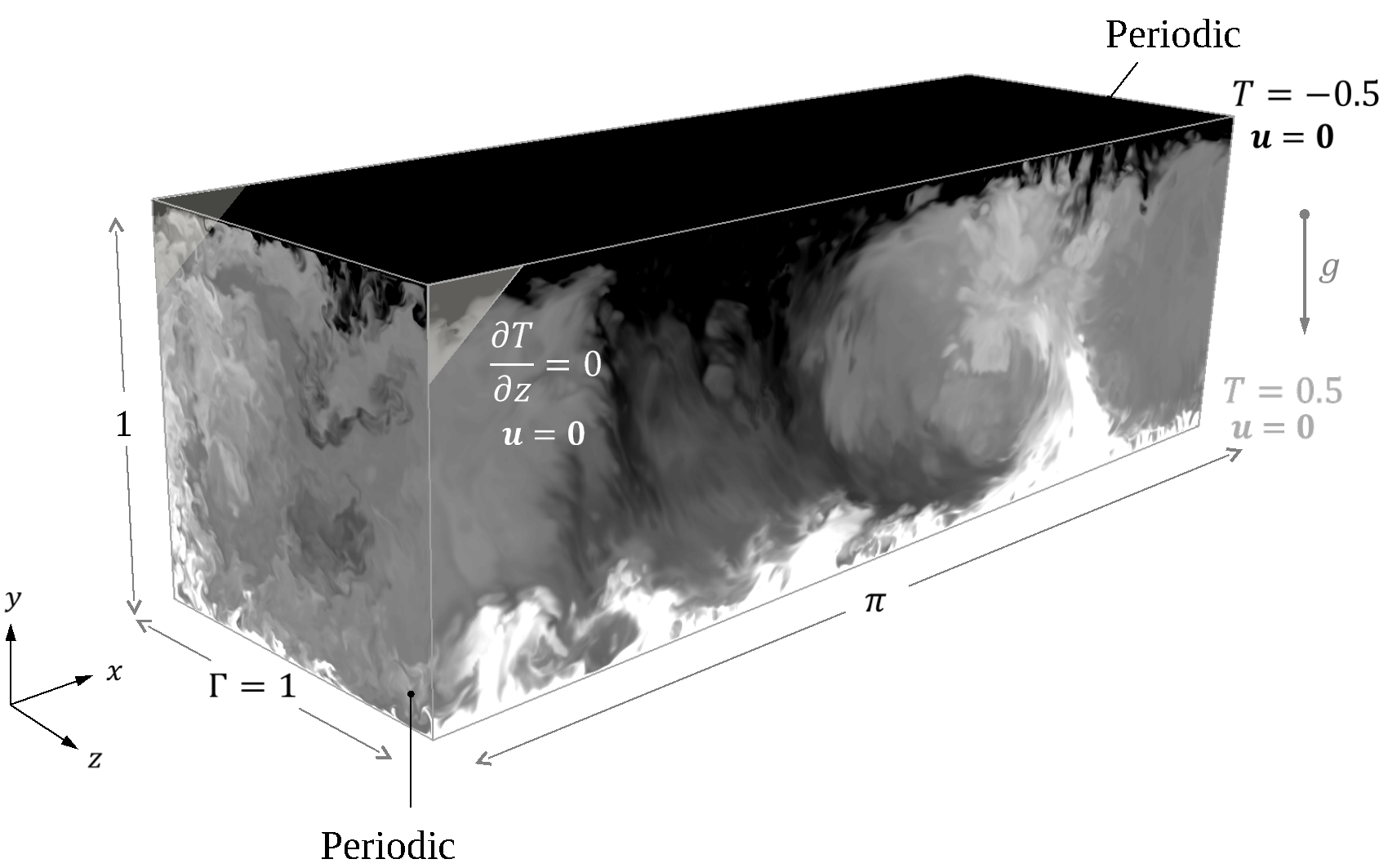}
\end{center}
\vspace{-1.69mm}
\begin{center}
    \includegraphics[width=\DNSflowswidth]{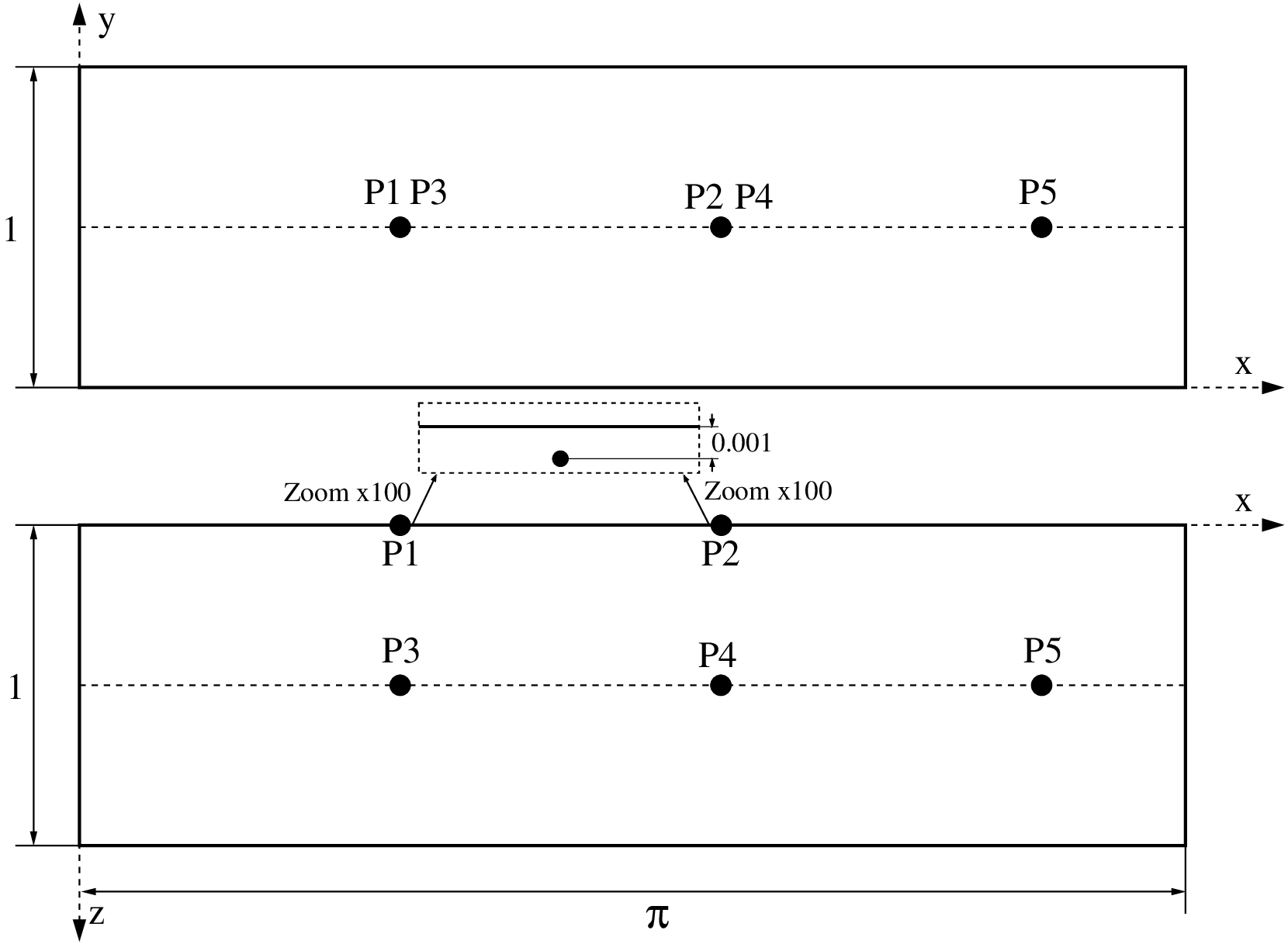}
\end{center}
\caption{\aft{Illustrative snapshot of the DNS simulation of
    air-filled ($Pr=0.7$) Rayleigh–B\'{e}nard configuration
    studied~\cite{DABTRI19-3DTOPO-RB}, together with a schematic
    representation of the computational domain and the location of the
    monitoring probes used in the analysis (see
    Table~\ref{points_RBC}). The simulations were performed at
    Rayleigh numbers up to $Ra=10^{11}$ on the MareNostrum~4
    supercomputer using $8192$~CPU-cores using a mesh with
    $5.7$~billion grid points.}}
\label{RBC}
\end{figure}

\begin{figure}[!t]
\begin{center}
\includegraphics[angle=-90,width=\largespectrawidth]{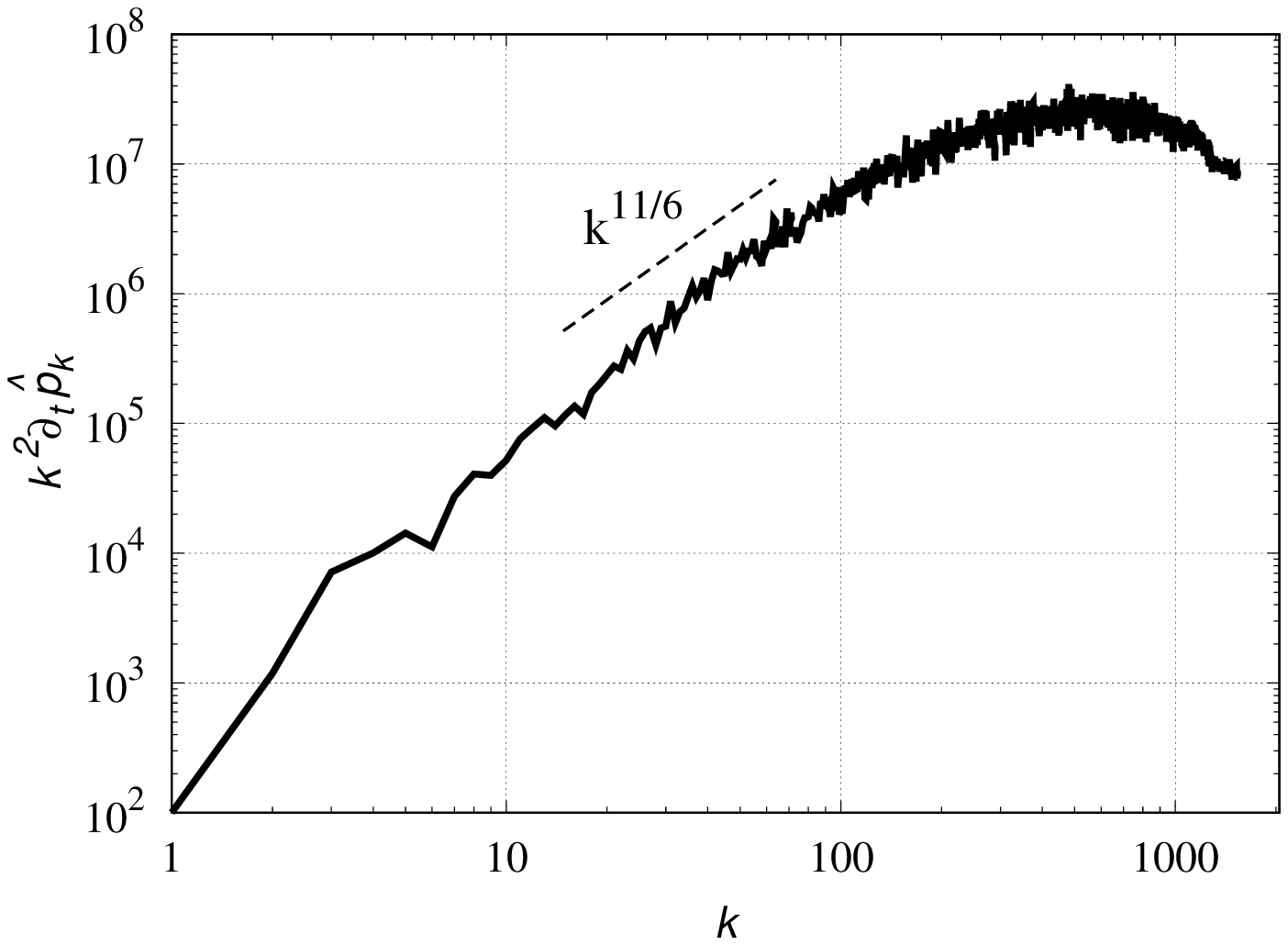}
\end{center}
\caption{Spectrum of the temporal derivative of the pressure field
  rescaled by $k^2$. Results correspond to the forced HIT simulation
  at $Re_\lambda \approx 433$ from the JHTDB
  database~\cite{PER07-JHTDB,LI08-JHTDB}}
\label{FFT_DtPres_HIT}
\end{figure}

\newcolumntype{x}[1]{>{\centering\arraybackslash\hspace{0pt}}p{#1}}

\begin{table}
\begin{center}
\begin{tabular}{x{10mm}|x{12mm}x{12mm}x{12mm}x{12mm}x{12mm}|x{12mm}x{12mm}x{12mm}x{12mm}}
        & \pSLa & \pSLb & \pSLc & \pSLd & \pSLe & \pWa & \pWb & \pWc & \pWd \\
\hline
$\x$    & $-0.45$ & $-0.40$ & $-0.30$ & $0.00$ & $0.50$ & $1.00$ & $1.50$ & $2.50$ & $3.50$ \\
$\y$    & $\hphantom{-}0.63$  & $\hphantom{-}0.63$  & $\hphantom{-}0.63$  & $0.63$ & $0.63$ & $0.00$ & $0.00$ & $0.00$ & $0.00$ \\
\end{tabular}
\end{center}
\caption{List of monitoring locations for the square
  cylinder. \XaviR{The first five probes, labeled \pSLa-\pSLe, are
    located in the shear-layer region, while the remaining probes,
    labeled \pWa-\pWd, are located in the wake region.}}
\label{points_CYL}
\end{table}

\begin{table}
\begin{center}
\begin{tabular}{x{10mm}|x{12mm}x{12mm}x{12mm}x{12mm}x{12mm}|x{12mm}x{12mm}x{12mm}x{12mm}}
        & \pPLa   & \pPLb   & \pPLc & \pPLd & \pPLe \\
\hline
$\x$    & $1$     & $2$     & $1$   & $2$   & $3$ \\
$\y$    & $0.5$   & $0.5$   & $0.5$ & $0.5$ & $0.5$ \\
$\z$    & $0.001$ & $0.001$ & $0.5$ & $0.5$ & $0.5$
\end{tabular}
\end{center}
\caption{List of monitoring locations for the Rayleigh--B\'{e}nard
  configuration. \XaviR{Probes \pPLa~and \pPLb~are located inside the
    boundary layer whereas probes \pPLc~to \pPLe~are in the bulk
    region}.}
\label{points_RBC}
\end{table}

\begin{figure}[!t]
\begin{center}
\includegraphics[angle=-90,width=\largespectrawidth]{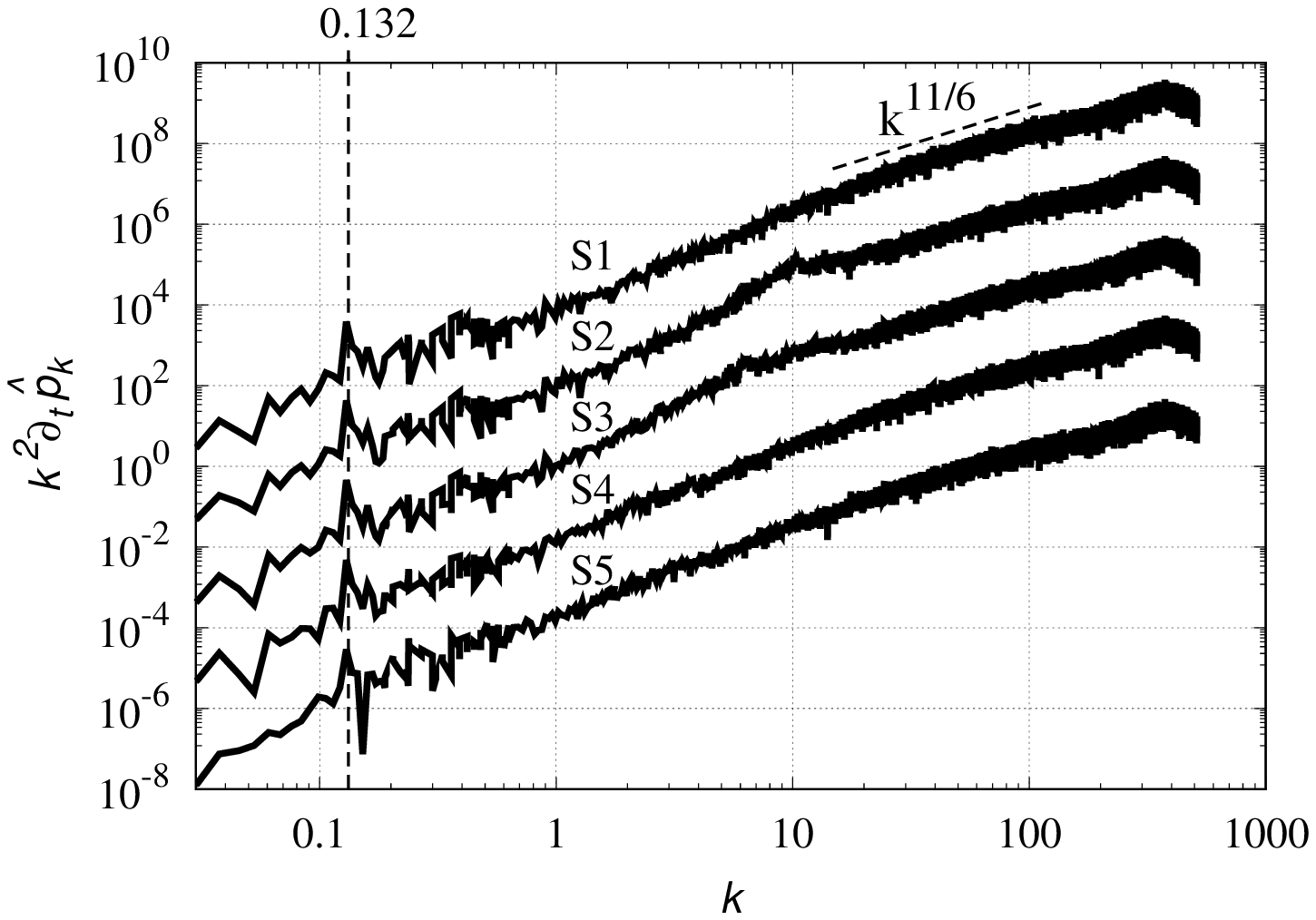}
\includegraphics[angle=-90,width=\largespectrawidth]{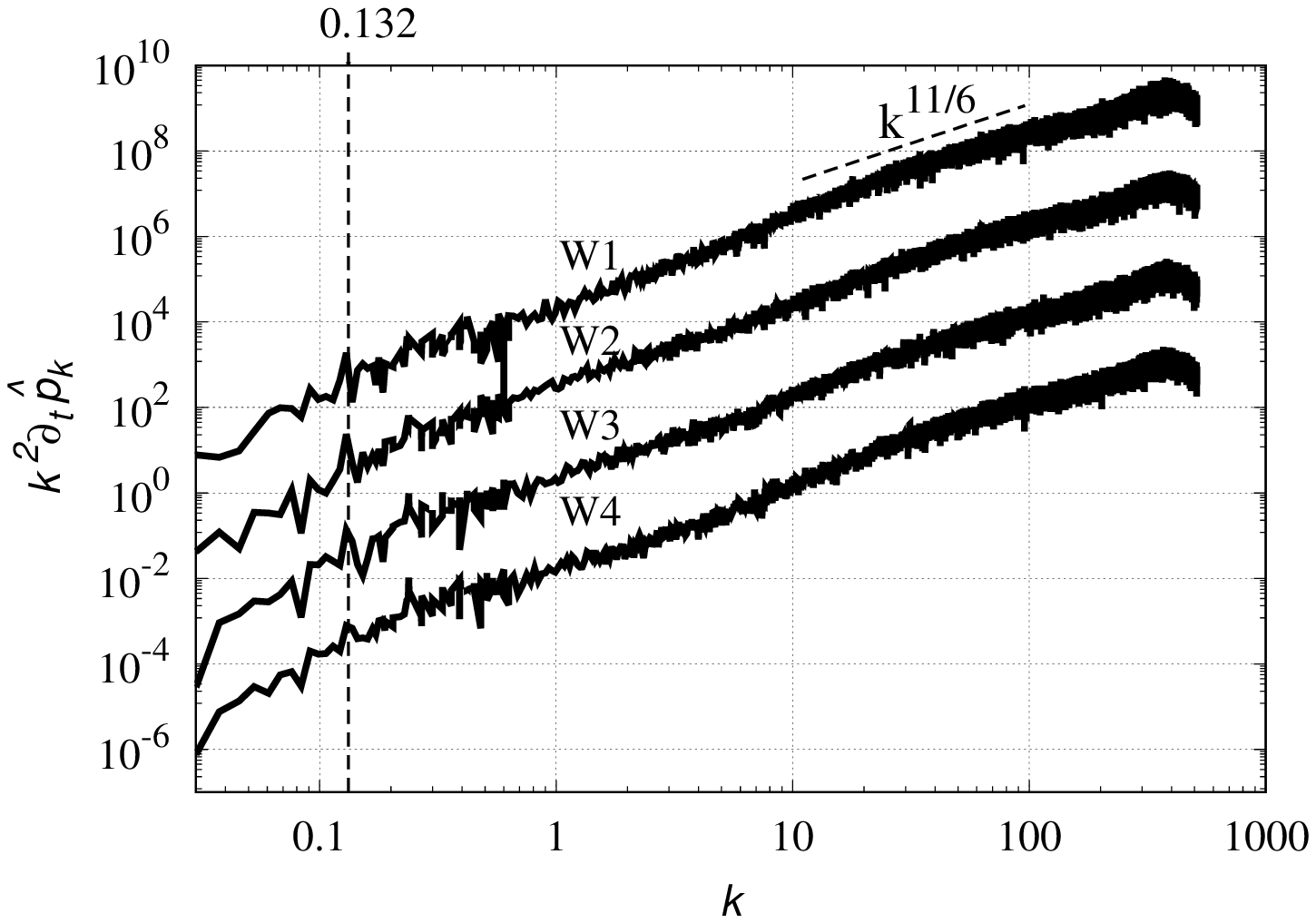}
\end{center}
\caption{Spectra of the temporal derivative of the pressure field
  rescaled by $k^2$. Results correspond to the turbulent flow around a
  square cylinder at $Re=22000$ displayed in Figure~\ref{CYL}. Top:
  results for a set of points located in the shear layer. Bottom:
  results for a set of points located in the wake regions. See
  Table~\ref{points_CYL}, for details.}
\label{FFT_DtPres_CYL22K}
\end{figure}

\begin{figure}[!t]
\begin{center}
\includegraphics[angle=-90,width=\largespectrawidth]{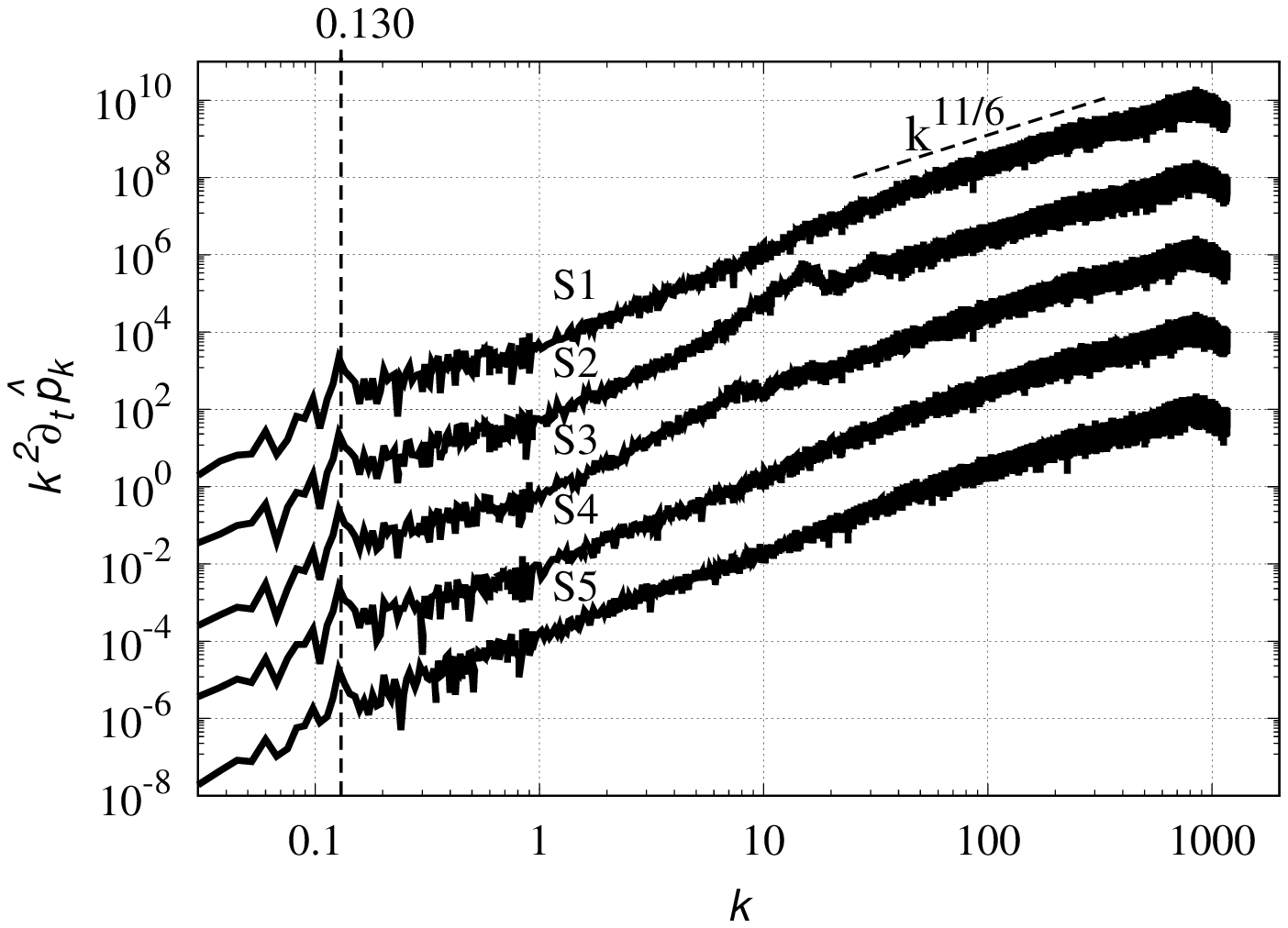}
\includegraphics[angle=-90,width=\largespectrawidth]{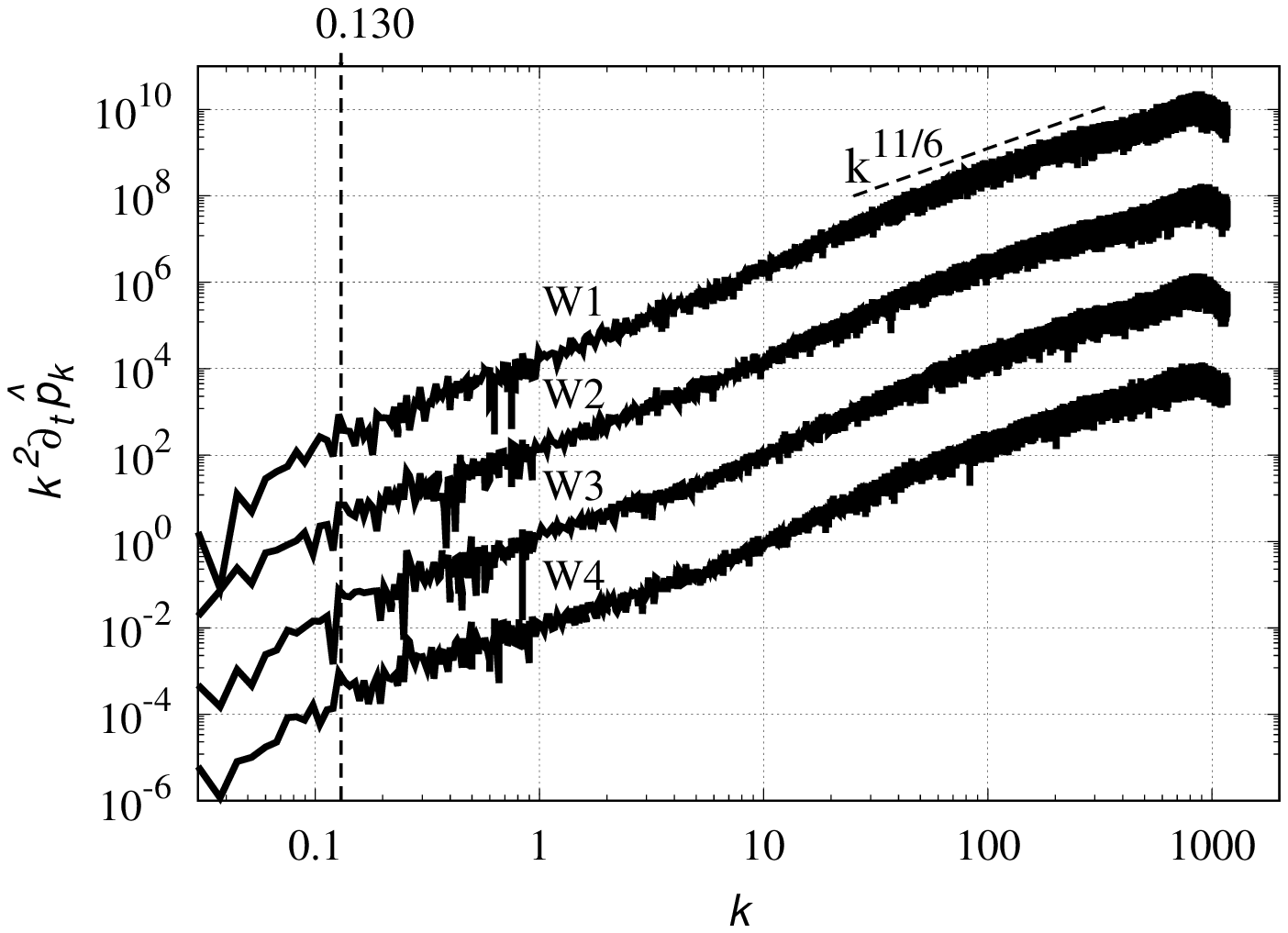}
\end{center}
\caption{Same as Figure~\ref{FFT_DtPres_CYL22K}, but for $Re=55000$.}
\label{FFT_DtPres_CYL55K}
\end{figure}

\begin{figure}[!t]
\begin{center}
\includegraphics[angle=-90,width=\largespectrawidth]{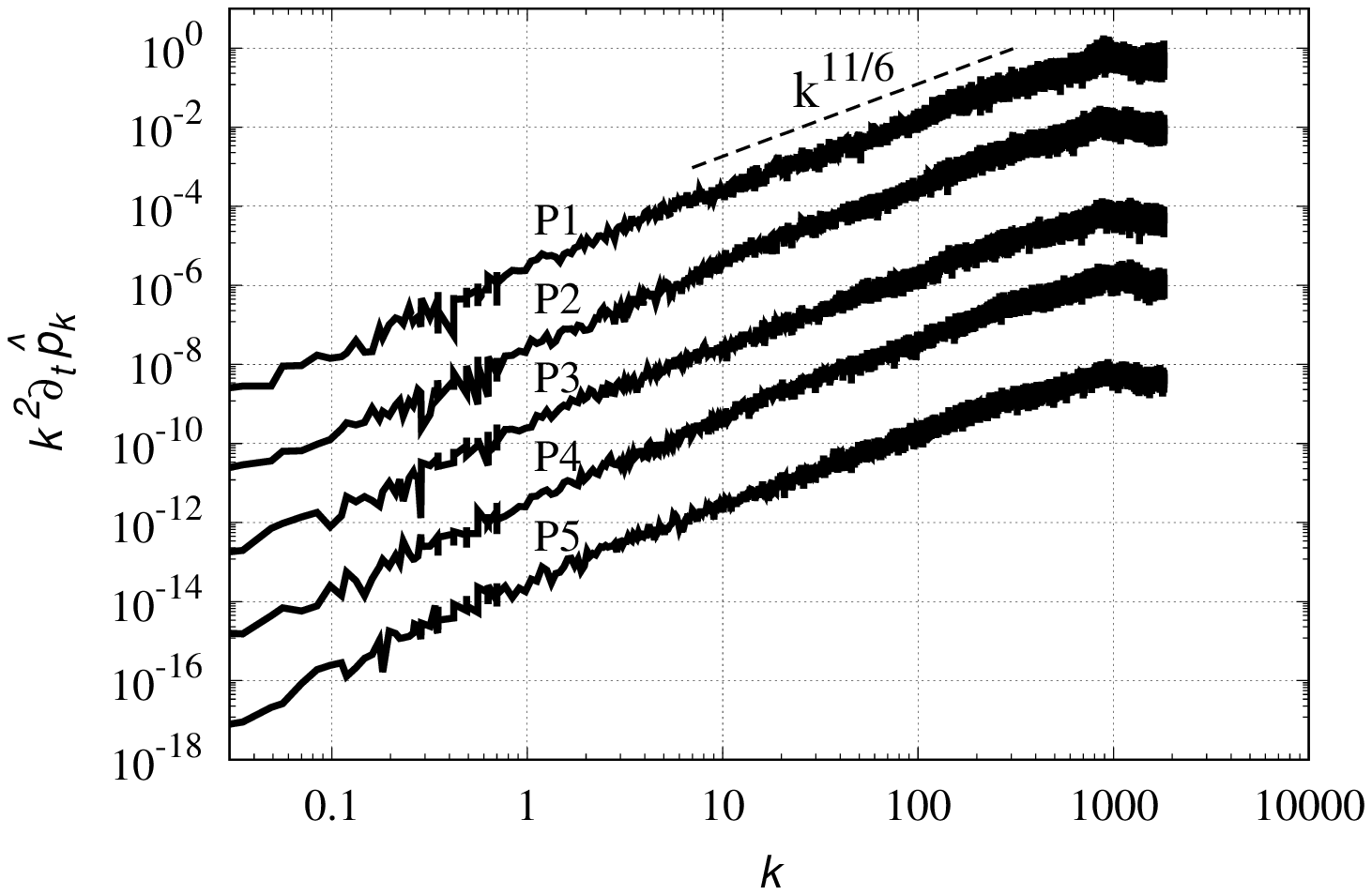}
\includegraphics[angle=-90,width=\largespectrawidth]{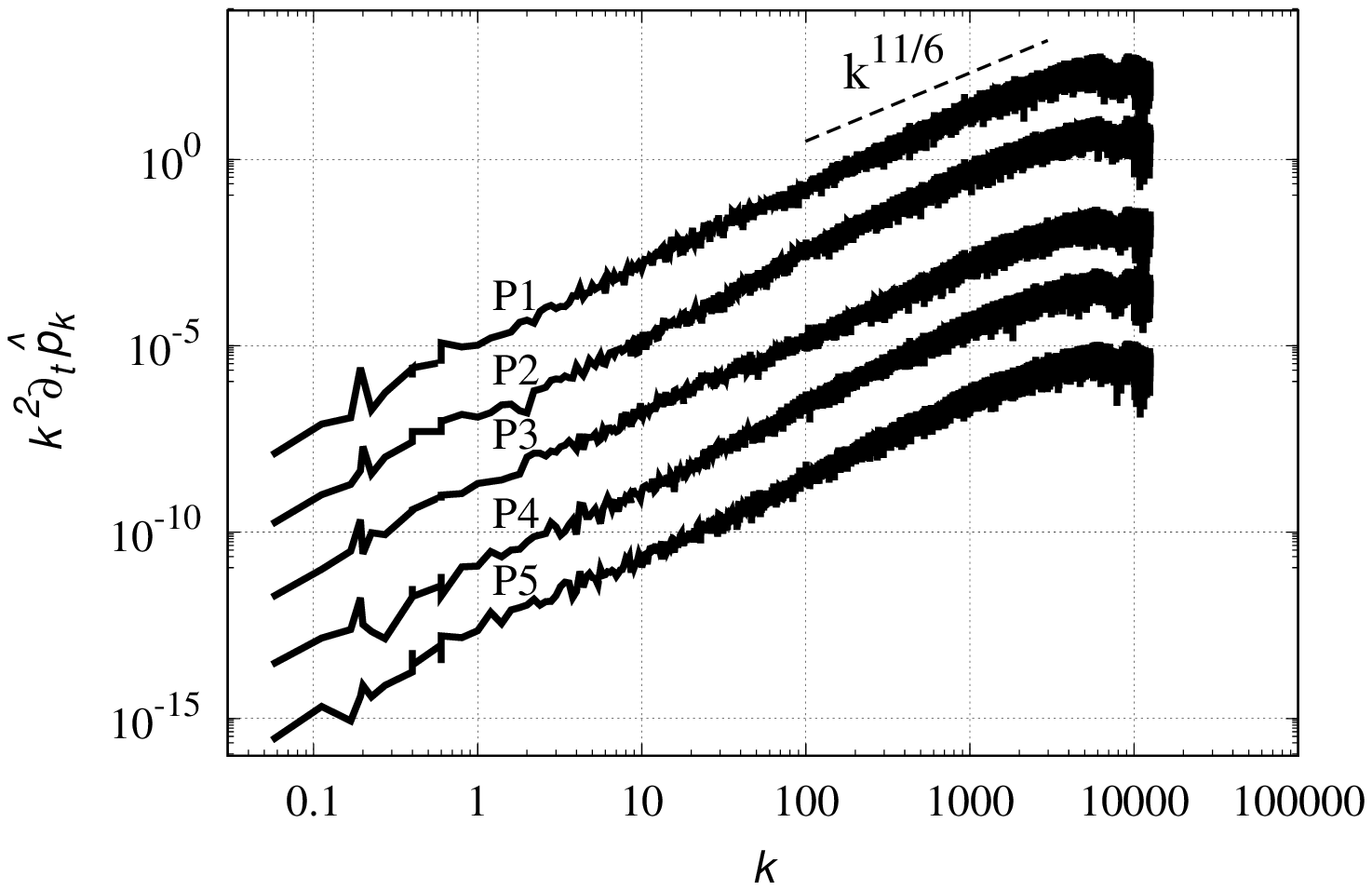}
\end{center}
\caption{Spectra of the temporal derivative of the pressure field
  rescaled by $k^2$. Results correspond to the air-filled
  Raleigh--B\'{e}nard convection flow at $Ra=10^{10}$ (top) and
  $10^{11}$ (bottom) displayed in Figure~\ref{RBC}. See
  Table~\ref{points_RBC}, for details.}
\label{FFT_DtPres_RBC}
\end{figure}

\subsection{Analysis of complex flows}

\label{complex_flows}

In this subsection, we examine two non-homogeneous turbulent flow
configurations: the flow around a square cylinder and an air-filled
Rayleigh–Bénard convection (RBC). Both cases involve strongly
inhomogeneous turbulence with a high degree of flow complexity. They
correspond to configurations investigated in previous
studies~\cite{DABTRI19-3DTOPO-RB,TRI14-CYL} and are illustrated in
\aft{Figure~\ref{RBC}}. Unlike HIT, the absence of homogeneity and
isotropy in these flows prevents the computation of \XaviR{a} fully
three-dimensional spatial spectra, so alternative analysis must be
employed. In this case, we can analyse the temporal evolution of the
pressure field at different relevant locations. Namely, recalling the
definition of the initial residual of the Poisson solver, $r^0$, given
in Eqs.(\ref{residual1}) and~(\ref{residual3}), we can relate the
temporal derivatives of the pressure field and the invariant $\QG$,
\begin{equation}
\lapl \press^{n} - \lapl \press^{n+1} \approx 2 \Dt \partial_t \QG \hphantom{kk} \Longrightarrow \hphantom{kk} \lapl \partial_t \press \approx 2 \partial_t \QG .
\end{equation}
\noindent leading to the following spectra relation and power-law
scaling
\begin{equation}
k^{2} \partial_t \hat{\press}_k \approx 2 \partial_t \hat{Q}_G \hphantom{kk} \stackrel{(\ref{res_k})}{\Longrightarrow} \hphantom{kk} k^{2} \partial_t \hat{\press}_k \propto k^{\scalResK} \hphantom{kk} \text{with} \hphantom{kk} \scalResK = 11/6 \hphantom{k} \text{(for NS)} .
\end{equation}
\noindent Nevertheless, this would still require computing a
shell-summed spectrum of $\partial_t \press$, which is not feasible
for non-homogeneous flows. Instead, we can analyze the corresponding
temporal spectra and invoke \XaviR{Taylor’s frozen-flow
  hypothesis~\cite{TAY38}} to relate them to \Xavi{the spectral
  distribution of the spatial scales}. To demonstrate the validity of
this approach, Figure~\ref{FFT_DtPres_HIT} presents the analysis for
the HIT case discussed in the previous subsection. For this dataset, a
sequence of $5028$~consecutive frames, stored every ten time-steps of
the DNS simulation, was available. To improve statistical convergence,
the temporal evolution of pressure was extracted at eight evenly
spaced locations, their individual spectra were computed, and the
results were subsequently averaged. The figure clearly shows the
expected $11/6$~scaling, thereby validating the proposed approach.

\mbigskip

At this stage, we used the same analysis for the two above-mentioned
configurations. Figures~\ref{FFT_DtPres_CYL22K}
and~\ref{FFT_DtPres_CYL55K} show results for the flow around a square
cylinder at $Re=22000$ and $Re=55000$, respectively. For the first
case, the numerical setup, including the mesh, discretization schemes,
and boundary conditions, was the same as in the original
study~\cite{TRI14-CYL}, \aft{which also provides a publicly available
  DNS database~\cite{PUBDAT}. A similar database for the $Re=55000$
  case is also made available in this work~\cite{PUBDAT55K}. These
  datasets are provided for the sake of reproducibility and as a
  reference for studies of turbulent flows and turbulence modeling
  around bluff bodies.} The grid resolution \aft{at $Re=22000$}
\bef{in this case} was $1272 \times 1174 \times 216$ in the
stream-wise, cross-stream and span-wise direction, respectively,
corresponding to approximately $323$~million grid points. For the
higher Reynolds number, the configuration was kept identical except
for the use of a finer \XaviR{mesh of $2544 \times 2348 \times 432
  \approx 2.6$~billion} grid points. For both Reynolds numbers, the
time evolution of the pressure was analyzed at the monitoring
locations listed in Table~\ref{points_CYL}. The first five probes,
labeled \XaviR{\pSLa-\pSLe}, are located in the shear-layer region,
while the remaining probes, labeled \XaviR{\pWa-\pWd}, are located in
the wake region. \Xavi{The same set of probes} were \Xavi{used} in our
previous study~\cite{TRI14-CYL} \Xavi{for characterizing the onset and
  development} of instabilities. The first five probes are placed near
the upper corner of the cylinder, where small vortices generated by
the Kelvin--Helmholtz instabilities rapidly develop and are convected
downstream. These structures are clearly visible in the instantaneous
snapshot of \aft{Figure~\ref{CYL}} (see also the corresponding
movie). As they evolve, the vortices grow in size and trigger the
transition to turbulence before reaching the downstream edge of the
cylinder. They eventually break up into finer scales and are entrained
by the much larger von K\'{a}rm\'{a}n vortices. The first
Kelvin--Helmholtz structure appears at $x \approx -0.45$
\XaviR{(point~\pSLa)}, in very good agreement with previous
experimental~\cite{BRU08a} and numerical
studies~\cite{TRI14-CYL}. However, the dominant frequency in both the
shear-layer and wake regions corresponds to the von~K\'{a}rm\'{a}n
mode, taking values of $0.132$ for $Re=22000$ (see
Figure~\ref{FFT_DtPres_CYL22K}) and \aftR{$0.130$} at $Re=55000$ (see
Figure~\ref{FFT_DtPres_CYL55K}), respectively. These results are in
excellent agreement with previous experimental observations and
confirm the very weak $Re$-number dependence of the Strouhal number in
this range of $Re$-numbers~\cite{BAI18}. \aftR{Consistently, the
  time-averaged drag coefficients obtained, ${C_D} \approx 2.18$ at
  $Re=22000$ and $C_D \approx 2.21$ at $Re=55000$, also agree very
  well with the reported asymptotic
  plateau~\cite{BAI18}}. Nevertheless, the most significant feature
observed in Figures~\ref{FFT_DtPres_CYL22K} 
and~\ref{FFT_DtPres_CYL55K} is the predicted $\scalResK=11/6$ scaling
at high wavenumbers, thereby confirming this power-law behavior for
non-homogeneous flows across different Reynolds numbers.

\mbigskip

As a second test-case for non-homogeneous flows, we consider an
air-filled ($Pr=0.7$) RBC at $Ra=10^{10}$ and $10^{11}$. These cases
were already investigated in a previous
study~\cite{DABTRI19-3DTOPO-RB}, where the flow topology and its main
features were analyzed in detail. Again, the numerical setup,
including the mesh, discretization schemes, and boundary conditions,
is the same as in the \Xavi{previous}
papers~\cite{DABTRI15-TOPO-RB,DABTRI19-3DTOPO-RB}. The \XaviR{mesh}
resolution is $1024 \times 768 \times 768 \approx 604$~million grid
points for $Ra = 10^{10}$, and $2048 \times 1662 \times 1662 \approx
5.7$~billion grid points in the homogeneous spanwise direction, the
horizontal cross stream direction and the vertical direction,
respectively. The flow exhibits strong inhomogeneity in the vertical
direction, with thin thermal and velocity boundary layers adjacent to
the horizontal isothermal plates and a plume-dominated bulk region. To
capture these distinct flow regions, five monitoring points are
considered in the analysis: namely, probes \pPLa~and \pPLb~are located
inside the boundary layer whereas probes \pPLc~to \pPLe~are in the
bulk \Xavi{region} (see Table~\ref{points_RBC}, for details). Despite
\Xavi{such} complexity, the spectra of $k^{2} \partial_t p$ showed in
Figure~\ref{FFT_DtPres_RBC} display very similar trends to those
observed for the square cylinder. Namely, both cases clearly show the
predicted $\scalResK = 11/6$ slope at high wavenumbers, thereby
confirming that the theoretical scaling extends robustly to
buoyancy-driven turbulence at very high Rayleigh numbers.

\mbigskip

The consistent results obtained for both the square cylinder and the
RBC at different $Re$ and $Ra$ numbers provide strong additional
support for the theoretical framework developed in this paper,
demonstrating its validity beyond the homogeneous case. They also
indicate that the solver convergence trends inferred from the theory
remain applicable in realistic CFD settings, where non-homogeneity and
geometric complexity are unavoidable. Nevertheless, the predicted
scaling of the number of solver iterations with respect to $Re$ (see
Eq.~\ref{xi_def}) still needs to be tested over a broader range of
Reynolds numbers. This issue is addressed in the following
\aft{subsections} \bef{subsection}.


%

\begin{figure}[!t]
\begin{center}
    \includegraphics[angle=-90,width=\largespectrawidth]{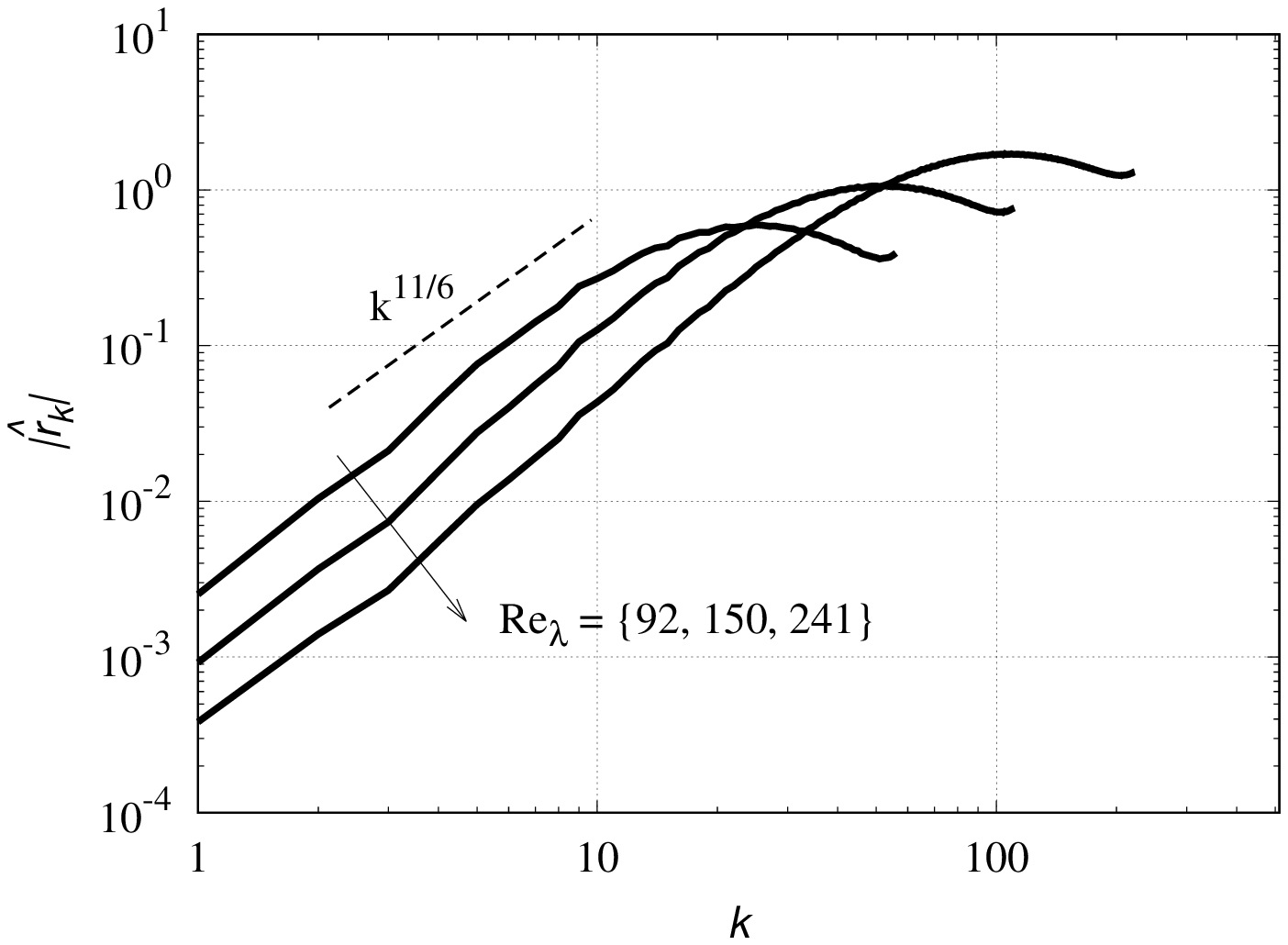}
    \includegraphics[angle=-90,width=\largespectrawidth]{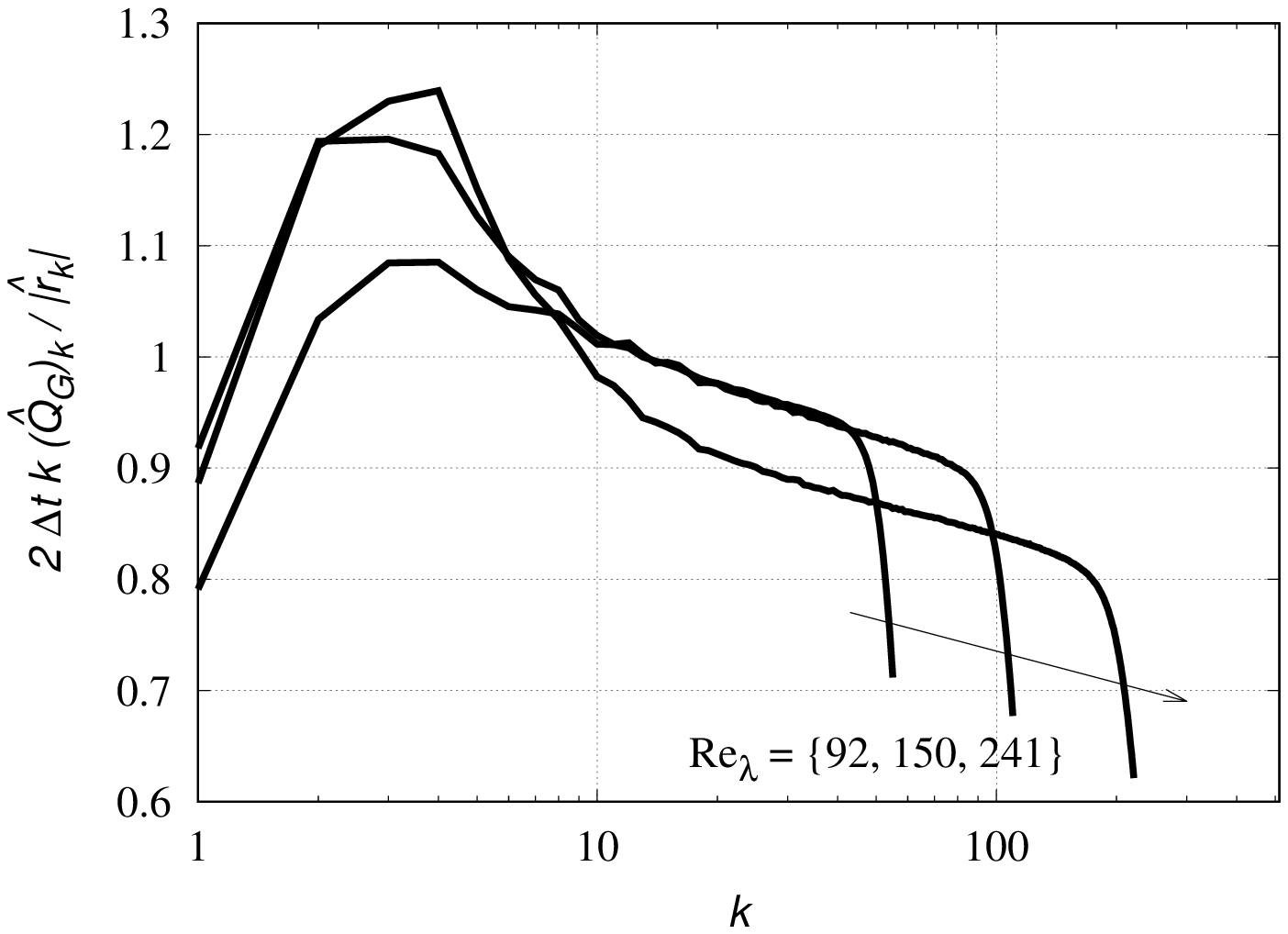}
\end{center}
\caption{\aft{Spectral analysis of the Poisson residual in forced HIT
    for different Reynolds numbers. Top: spectrum of the initial
    residual, showing the predicted $\hat{r}_k \propto k^{11/6}$
    scaling. Bottom: comparison between the residual spectrum and the
    spectrum of the second invariant $\QG$, illustrating the relation
    $\hat{r}_{k} \approx 2 \Dt k (\hat{Q}_{G})_{k}$ over the inertial
    range.}}
\label{HIT_residual_spectra}
\end{figure}

\begin{figure}[!t]
\begin{center}
    \includegraphics[angle=-90,width=\largespectrawidth]{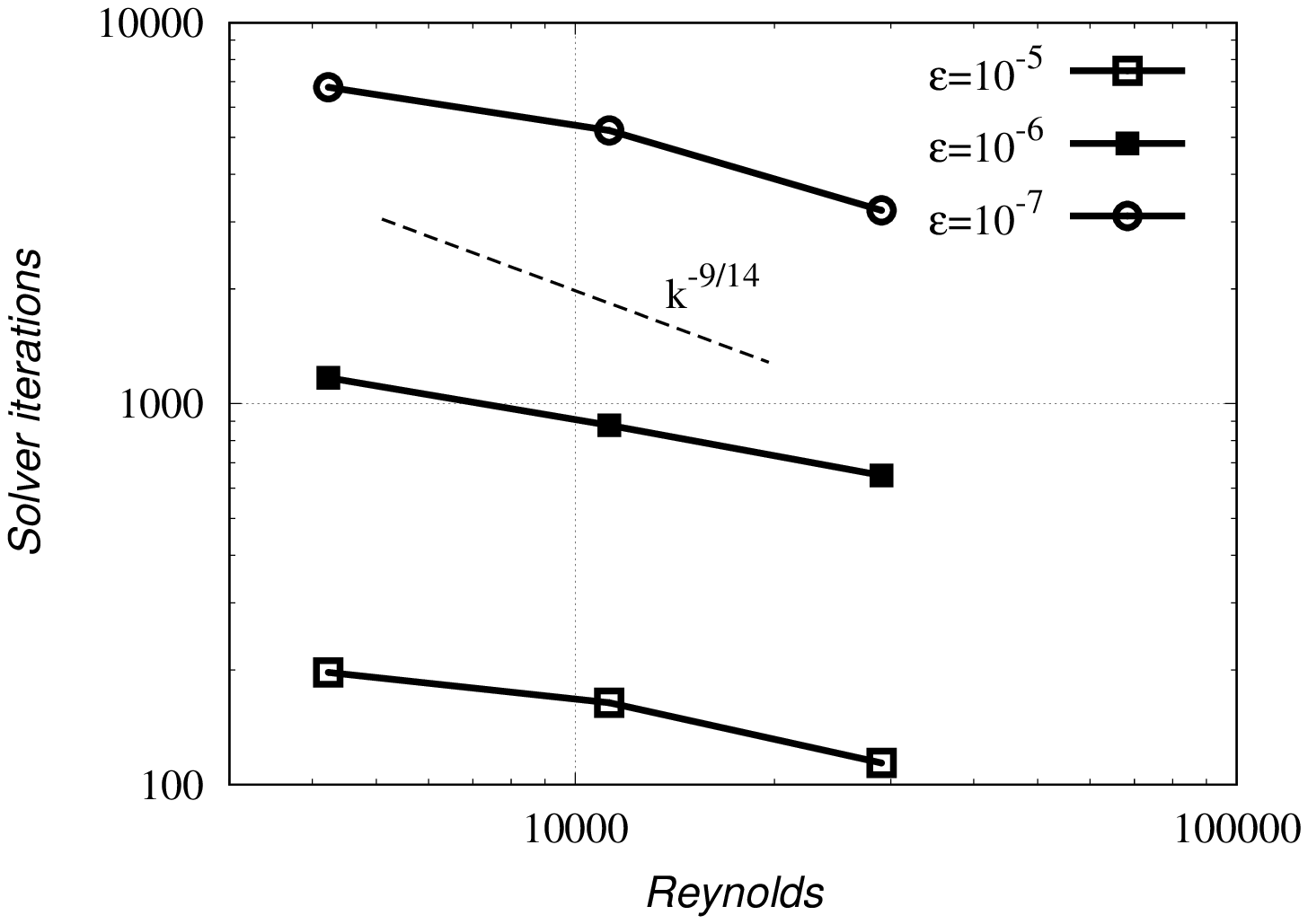}
\end{center}
\caption{\aft{Scaling of the number of Jacobi iterations required to
    reach a given residual threshold in forced HIT as a function of
    the Reynolds number.}} 
\label{HIT_solver_scaling}
\end{figure}

\subsection{\aft{Reynolds-number scaling in homogeneous isotropic turbulence}}

\label{HIT_tests}

\aft{In this subsection, we analyse the Reynolds-number scaling of
  both the residual of the Poisson equation and the convergence
  behaviour of the solver. This could not be addressed in the previous
  test cases due to the limited range of Reynolds numbers considered.
  To overcome this limitation, additional simulations of forced HIT
  have been performed, providing a controlled setting to investigate
  these effects. Simulations were carried out at resolutions of
  $128^3$, $256^3$, and $512^3$ grid points, corresponding to
  Taylor-scale Reynolds numbers of approximately $Re_\lambda \approx
  92$, $150$, and $241$, respectively. All simulations were performed
  using the open-source SpNS code~\cite{SpNS-yotta}, a pseudo-spectral
  incompressible NS solver employing a classical $3/2$~dealiasing rule
  and a second-order fully explicit time-integration scheme with a
  self-adaptive strategy~\cite{TRI22-AlgEigCD}. The forcing is applied
  at the largest scale by maintaining the energy content of the lowest
  wavenumber modes constant in time.}

\mbigskip

\aft{Figure~\ref{HIT_residual_spectra} (top) shows the spectral
  distribution of the initial residual of the Poisson equation for the
  different HIT simulations. The residual is non-dimensional, using
  the root-mean-square velocity of the largest scale as reference
  velocity and the domain size as reference length. In all cases, the
  residual exhibits the predicted scaling $\hat{r}_k \propto k^{11/6}$
  over a broad range of wavenumbers. Deviations are observed only at
  the highest wavenumbers, where viscous effects are expected to
  dominate.  To further investigate the origin of this scaling, the
  residual spectra are compared with the corresponding spectra of the
  second invariant $\QG$. As shown in
  Figure~\ref{HIT_residual_spectra} (bottom), the relation $\hat{r}_k
  \approx 2 \Dt k (\hat{Q}_G)_{k}$ is observed to hold over a wide
  range of scales for all the Reynolds numbers considered. This
  provides strong support for both the idea that the convective
  contribution, $( \vel \cdot \nabla) \QG$, governs the spectral
  behaviour of the residual in the inertial range, and for the scaling
  assumption used in Eq.(\ref{convQscaling}), which relies on the
  Taylor's frozen-turbulence hypothesis.}

\mbigskip

\aft{Finally, the convergence behaviour of the Poisson solver has been
  analysed using a Jacobi solver. Figure~\ref{HIT_solver_scaling}
  shows the evolution of the number of iterations required to reach a
  given residual threshold as a function of the Reynolds number, which
  is related to the Taylor-scale Reynolds number, $Re_\lambda$, as $Re
  \approx \tfrac{1}{2} Re_{\lambda}^{2}$ (see Section~6.5.7 of
  Pope~\cite{POP00}). As expected, the number of iterations increases
  as the target residual, $\varepsilon$, is reduced. The observed
  trend is broadly consistent with the theoretical predictions, with a
  tendency towards the expected scaling as the Reynolds number
  increases and for stricter convergence criteria. However, the
  agreement with the predicted $Re^{-9/14}$ scaling is not yet fully
  satisfied, likely due to the relatively low Reynolds numbers
  considered and the limited extent of the inertial
  range. Nevertheless, the trend clearly improves for the highest
  Reynolds number investigated, supporting the validity of the
  proposed framework. It is worth noting that, although the range of
  Reynolds numbers covered in these simulations is wider than in the
  previous test cases, it remains relatively limited due to the
  prohibitive computational cost of fully resolved three-dimensional
  NS simulations. Nevertheless, these results provide a consistent
  picture linking the spectral properties of the flow to the
  convergence behaviour of the Poisson solver, supporting the
  theoretical framework proposed in this work in a fully
  incompressible NS setting. A numerical study over a much wider
  effective Reynolds-number range is presented in the next
  subsection.}


\subsection{Towards very high Reynolds numbers}

\label{Burgers}

To further assess the validity of the theoretical framework at very
high Reynolds numbers, we consider \Xavi{the} Burgers’ equation
\Xavi{in a periodic one-dimensional domain} as a simplified model
problem. \aft{In non-dimensional form it reads}
\begin{equation}
\frac{\partial u}{\partial t} + u \frac{\partial u}{\partial \x} = \frac{1}{Re} \frac{\partial^2 u}{\partial x^2} + f .
\end{equation}
The equation is solved using a pseudo-spectral approach with the
standard $3/2$ dealiasing rule applied to the non-linear convective
term. The forcing term, $f$, acts only at the smallest wavenumber,
$k=1$, keeping \Xavi{its} energy constant to unity, \ie~$E_{1}=1$. The
simulations are advanced in time until a statistically steady state is
reached. Once convergence is achieved, the resulting velocity field is
projected onto the space of divergence-free functions, which in this
simplified setting reduces to solving a one-dimensional Poisson
equation by means of either a Jacobi iterative solver or a MG solver
using Jacobi as smoother at each level. \aft{Therefore, these
  simulations of the Burgers' equation should be interpreted as a
  controlled numerical experiment designed to test the scaling
  behaviour of the Poisson solver rather than as a physical model of
  incompressible turbulence.} The analysis covers a very wide range of
$Re$-numbers from $Re=2^{5}=32$ up to $Re=2^{21} \approx 2.1M$. They
are solved with $N=4 Re$ Fourier modes, \ie~from $N=2^{7}=128$ up to
$N=2^{23} \approx 8.4M$. This linear resolution criterion arise from
the fact that, according to the classical Cole--Hopf
transformation~\cite{COL51,HOP50}, the smallest dissipative scale in
the 1D Burgers' equation is inversely proportional to the Reynolds
number. The adopted resolution is, in practice, very similar to that
recommended in recent studies \cite{LUO25}. Hence, for the 1D Burgers'
equation, $\Dx \sim l Re^{-1}$, where $l$ is the characteristic length
scale of the largest flow structures. \aft{To facilitate
  reproducibility and further analysis, the datasets generated for the
  Burgers' equation simulations used in this work are publicly
  available\cite{PUBDAT_Burgers}. The database contains spectral data
  obtained from DNS of the 1D Burgers' equation, providing a reference
  dataset for studies of extreme-Re scaling in Burgers turbulence.}

\mbigskip

Then, following the same arguments as in Eqs.(\ref{CFL_based_conv})
and (\ref{CFL_based_diff}), it leads to
\begin{equation}
\label{Dt_scaling_Burgers}
\frac{\Dt}{t_l} \sim \frac{1}{N_t} \sim Re^{\scalDtRey} \hphantom{kk} \text{with} \hphantom{kk}  \scalDtRey = -1 \hphantom{kkk} \text{(for Burgers' equation)} \XaviR{,}
\end{equation}
\noindent which is the counterpart of Eq.(\ref{Dt_scaling}). Notice
that, in this case, the Reynolds-number scaling is the same whether
the CFL stability constraint (see Eq.~\ref{CFL}) is limited by
convection or by diffusion.

\mbigskip

\begin{figure}[!t]
\begin{center}
\includegraphics[angle=-90,width=\largespectrawidth]{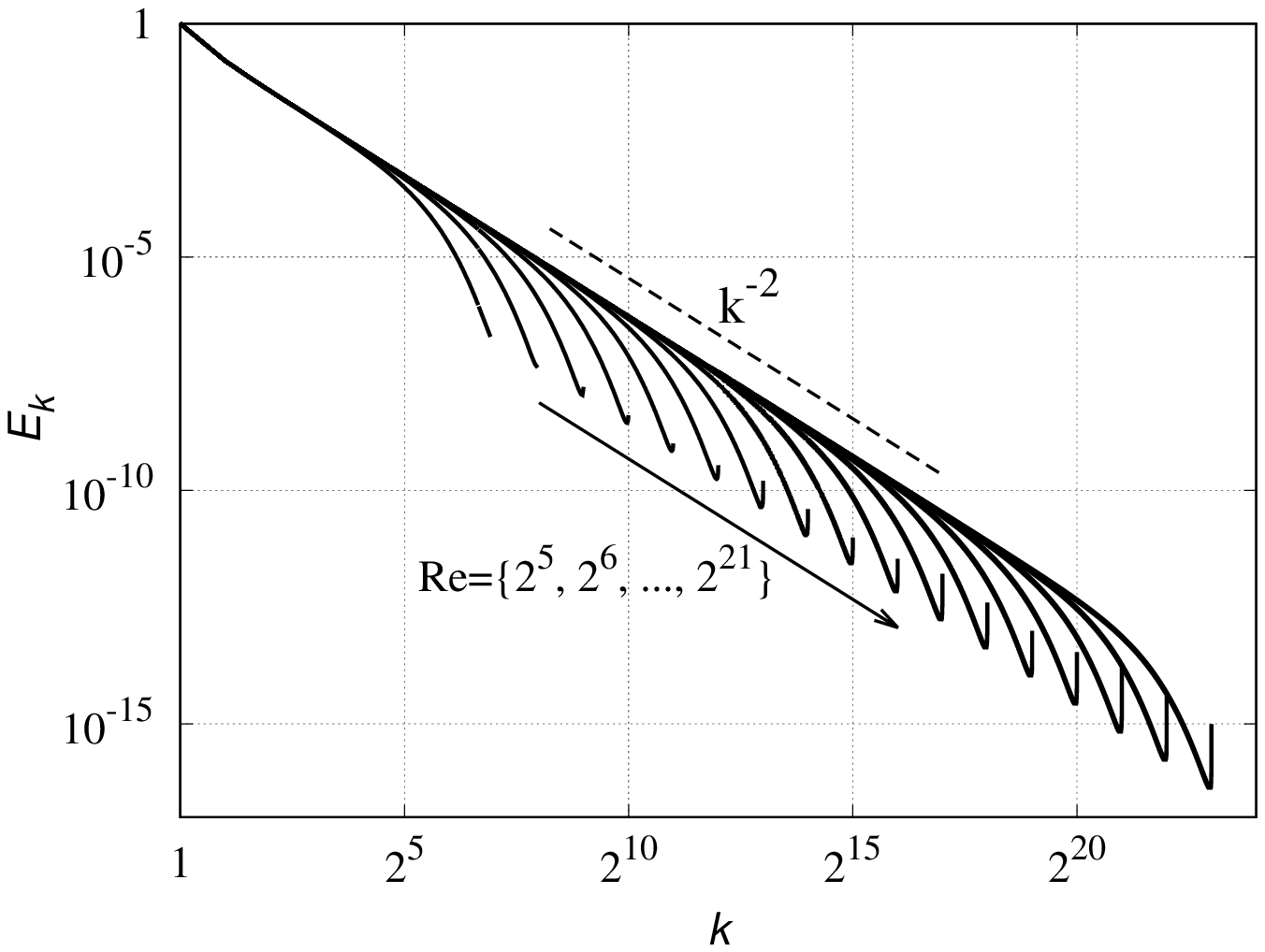}
\end{center}
\caption{Energy spectra for the forced Burgers' equation for
  $Re=\{2^5, 2^6, \dots, 2^{21}\}$, which have been numerically solved
  using $N=\{2^{7}, 2^{8}, \dots 2^{23}\}$ Fourier modes.}
\label{Burgers_energy_spectra}
\end{figure}

Apart from this, \Xavi{the} theoretical arguments developed in
Section~\ref{residual} have to be adapted to the scaling properties of
the Burgers’ equation (see Table~\ref{scalings}). Namely, in this
case, the slope for the \Xavi{solver's} residual is $\scalResK =
3$. This follows for the well-known $k^{-2}$ energy
spectrum~\cite{ALA22}, which can be clearly observed in the spectra
shown in Figure~\ref{Burgers_energy_spectra}. Namely, \Xavi{applying}
the same arguments used in Section~\ref{Re_scalings}, it leads to the
following relation for the residual
\begin{equation}
r^0 \approx \Dt^{\scalResDt} \partial_t \partial_x ( u \partial_x u ) \hphantom{kk} \text{with} \hphantom{kk}  \scalResDt=\left\{ \begin{array}{ll} 1 & \text{if $r^{0}$ defined as \Xavi{in} Eq.(\ref{residual1})} \\ 2 & \text{if $r^{0}$ defined as \Xavi{in} Eq.(\ref{residual2})} \end{array} \right.
\end{equation}
\noindent which is the counterpart of Eq.(\ref{residual3}). Then, we
can easily relate the $k^{-2}$ scaling in kinetic energy with the
scaling of the convective term, $ u \partial_x u$, using the
equilibrium hypothesis (see Eq.~\ref{equilibrium}),
\begin{equation}
( \widehat{u \partial_x u} )_k \sim \frac{k^{2}}{Re} \hat{u}_k \propto k^{2} k^{-1} = k .
\end{equation}
\noindent Finally, following the same line of arguments as in
Eqs.(\ref{convQscaling}) and~(\ref{res_k3}) leads to 
\begin{equation}
\label{res_k2_Burgers}
\boxed{\hat{r}_{k}^{0} \propto Re^{-1} \Dt^{\scalResDt} k^{\scalResK}  \hphantom{kk} \text{with} \hphantom{kk} \scalResK = 3 \hphantom{kk} \text{and} \hphantom{kk} \scalResDt=\left\{ \begin{array}{lc} 1 & \text{if $\hat{r}$ defined as \Xavi{in} Eq.(\ref{residual1})} \\ 2 & \text{if $\hat{r}$ defined as \Xavi{in}Eq.(\ref{residual2})} \end{array} \right.}
\end{equation}
\noindent which is the counterpart of Eq.(\ref{res_k2}).

\mbigskip

\begin{figure}[!t]
\begin{center}
\includegraphics[angle=-90,width=\spectrawidth]{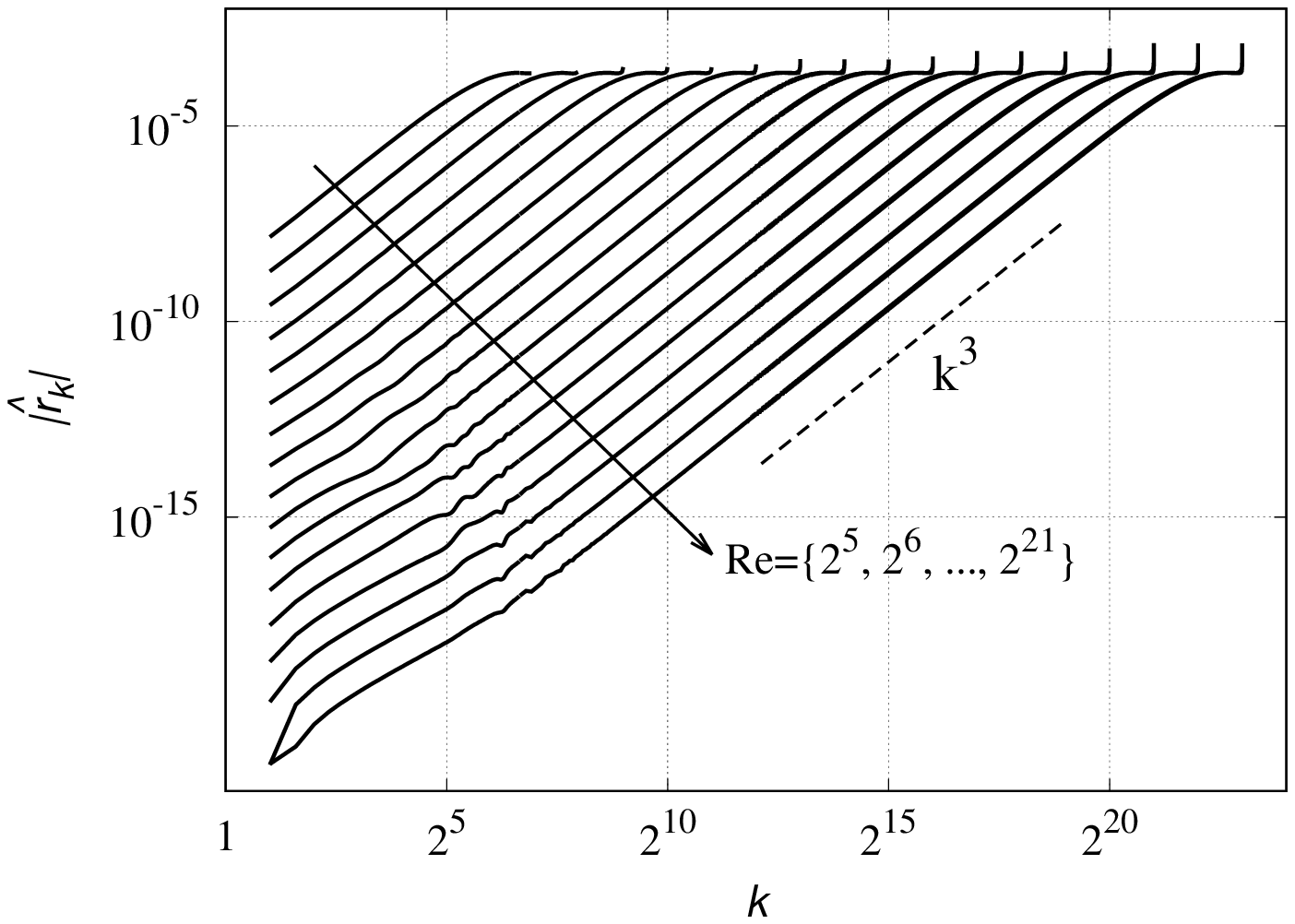}
\includegraphics[angle=-90,width=\spectrawidth]{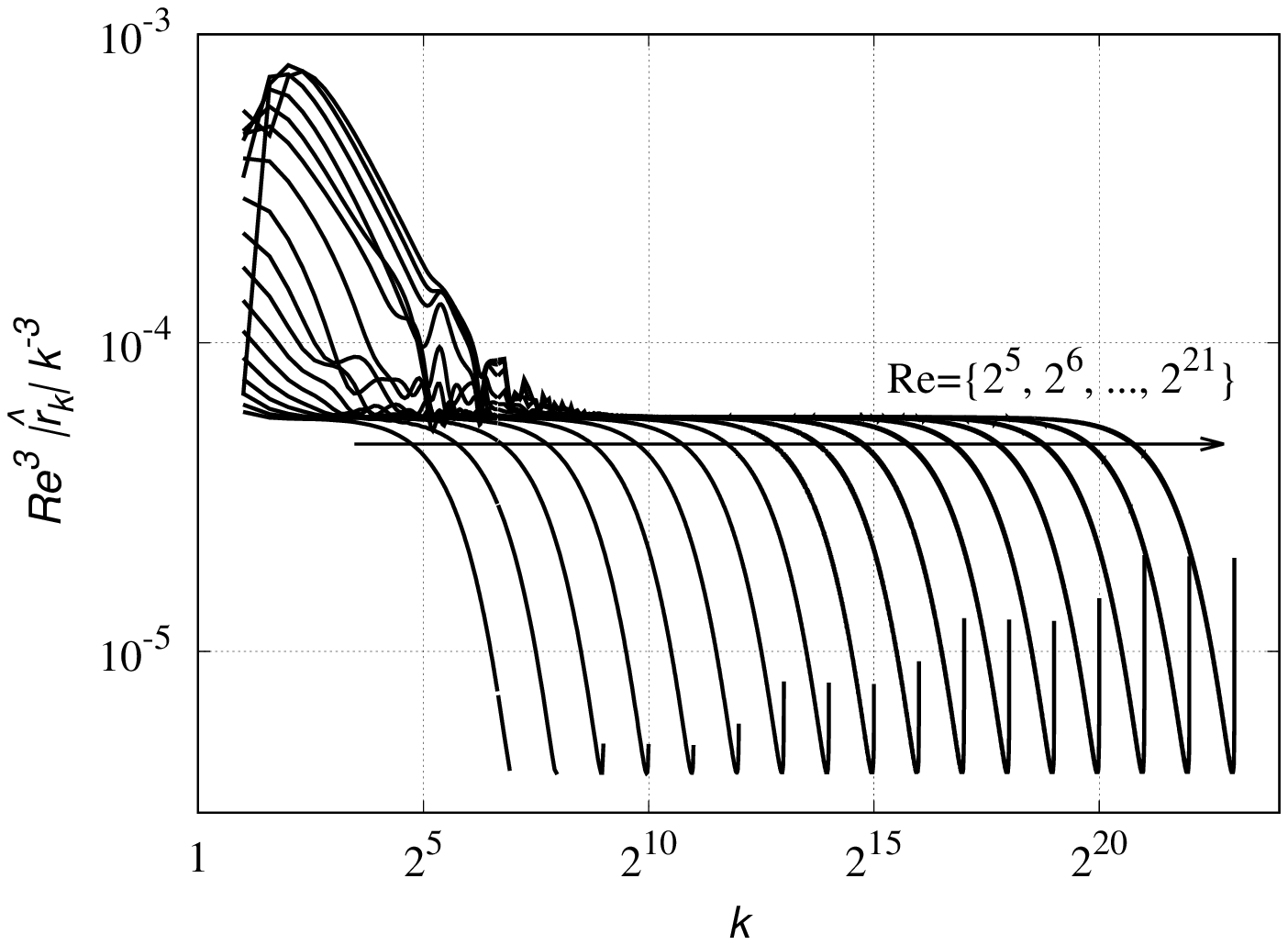}
\end{center}
\caption{Same as in Figure~\ref{Burgers_energy_spectra}, but for the
  residual of the Poisson equation (left) and its compensated spectra
  (right).}
\label{Burgers_residual_spectra}
\end{figure}

Results shown in Figure~\ref{Burgers_residual_spectra} (left) support
the predicted $k^{3}$ scaling of the initial residual,
$\hat{r}^{0}_{k}$. Moreover, the compensated spectra in
Figure~\ref{Burgers_residual_spectra} (right) demonstrate that all
curves collapse irrespective of the Reynolds number, confirming the
validity of the scaling law given in Eq.~(\ref{res_k2_Burgers}). In
this particular case, $q=2$, \Xavi{corresponds} to the definition of
the residual in Eq.~(\ref{residual2}), and $\Delta t \sim Re^{-1}$ as
given in Eq.~(\ref{Dt_scaling_Burgers}), which together yield the
overall $Re^{-3}$ dependence observed in
Figure~\ref{Burgers_residual_spectra}. Note that the discrepancies at
very low wavenumbers in Figure~\ref{Burgers_residual_spectra} (right)
arise from the amplification introduced by the \XaviR{$k^{-3}$ scaling
  factor}.

\begin{figure}[!t]
\begin{center}
\includegraphics[angle=-90,width=\largespectrawidth]{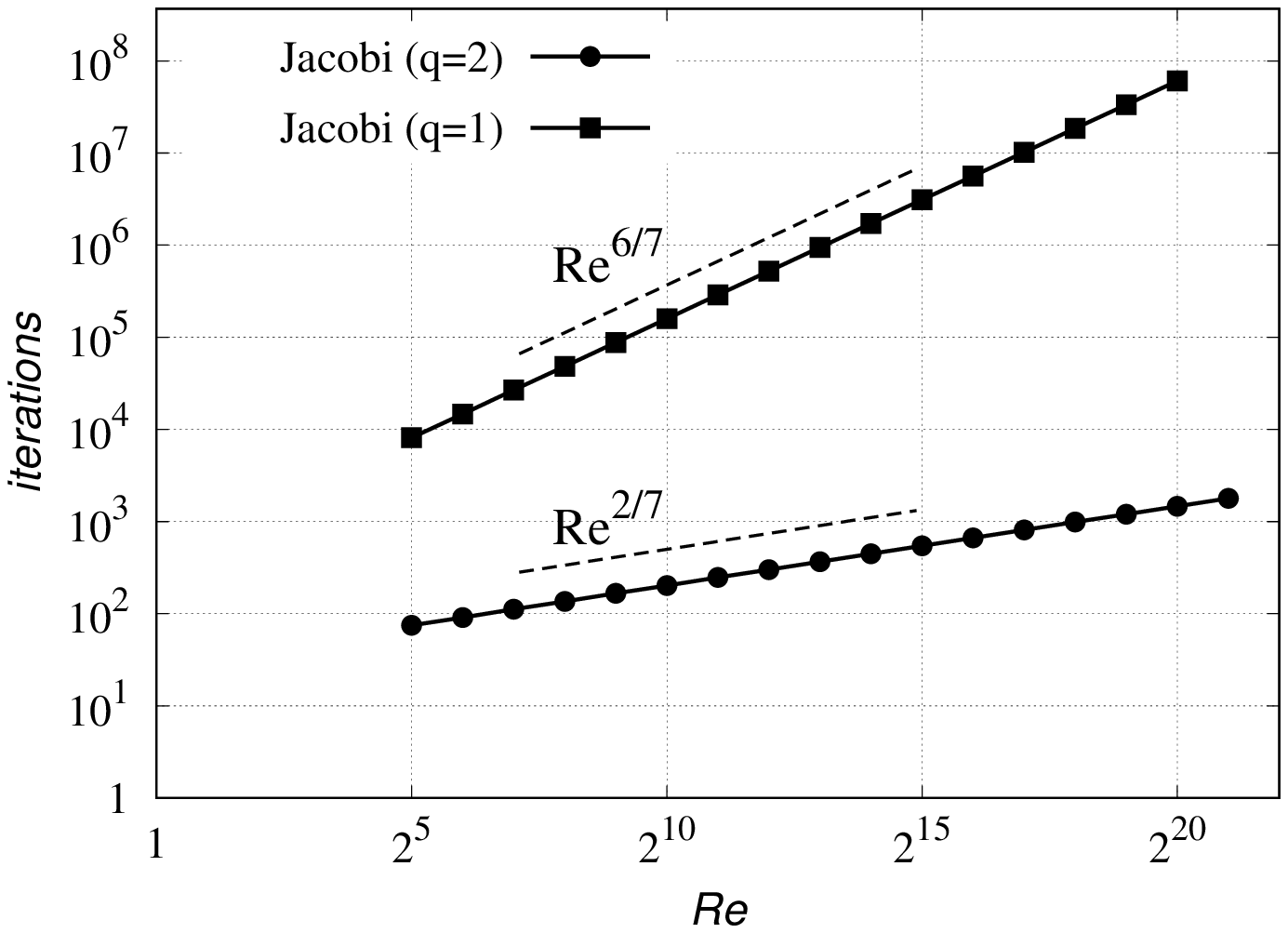}
\includegraphics[angle=-90,width=\largespectrawidth]{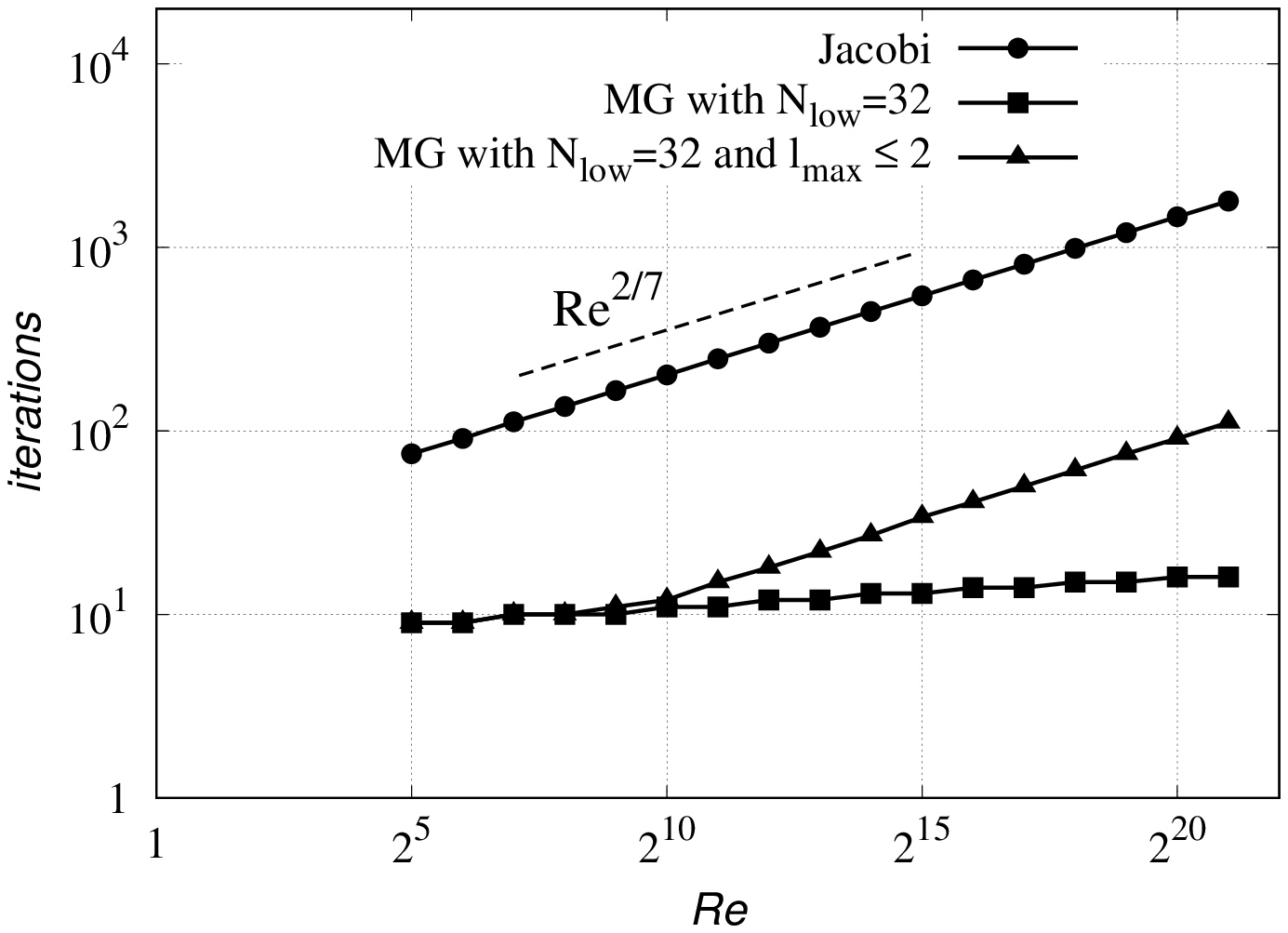}
\end{center}
\caption{Same as in Figure~\ref{Burgers_energy_spectra}, but for the
  total number of iterations required to solve the Poisson equation
  for different solver options. Top: Jacobi solver with
  $\scalResDt=\{1,2\}$. Bottom: comparison between Jacobi and MG
  solver with $\scalResDt=2$.}
\label{Burgers_iterations}
\end{figure}

The final analysis shown in Figure~\ref{Burgers_iterations} focuses on
the dependence of the solver iteration count on the Reynolds
number. As discussed above, the Poisson equation is solved using both
a Jacobi solver and a MG solver \Xavi{employing} Jacobi as a smoother
at each level. The latter corresponds to the analysis done in
Section~\ref{MG_solver}. In all cases, the convergence criterion is
set to $||r|| \le 10^{-6}$. \aft{Since all variables are expressed in
  non-dimensional form, the residual is also non-dimensional, and the
  same tolerance is applied consistently across all simulations.}
Results obtained within this wide range of Reynolds numbers using the
Jacobi solver are displayed in Figure~\ref{Burgers_iterations}
(top). They exhibit excellent agreement with the predicted scalings
(see the last column of Table~\ref{scalings}), namely $\scalnRey=2/7$
for $\scalResDt=2$ and $\scalnRey=6/7$ for $\scalResDt=1$. This
confirms that the proposed framework remains valid not only for
canonical turbulent flows such as HIT, RBC, or bluff-body wakes, but
also for this simplified model at very high Reynolds
numbers. Furthermore, results obtained with the Jacobi solver for
$\scalResDt=2$ are compared with the MG solver considering two
scenerios: (i) fixing the size of the coarsest MG level to
$N_{\text{low}}=2^{5}=32$ modes, \ie~$l_{\max} = (\log_{2} N - 1)/5$,
and (ii) using the same configuration but limiting $l_{\max} \le 2$,
\ie~$l_{\max} = \min{(\log_{2} N - 1)/5, 2}$. Thus, both
configurations coincide for small $N$, whereas for large values of
$N$, the latter recovers the $\scalnRey = 2/7$ scaling.

\section{Concluding remarks}

\label{conclusions}

In this work, we have combined physical reasoning and numerical
analysis to examine how the computational cost of solving the pressure
Poisson equation evolves with increasing Reynolds number in
simulations of incompressible flows. By analyzing the spectral
distribution of the solver residual, two competing mechanisms were
identified: the reduction of the time step at higher Reynolds numbers,
which improves the quality of the initial guess, and the refinement of
the computational mesh, which worsens the conditioning of the discrete
operator. The balance between these effects determines whether the
convergence of the solver accelerates or deteriorates as $Re$-number
increases.

\mbigskip

For NS turbulence, our theoretical analysis predicts that the
beneficial effect of smaller time steps dominates. Consequently, the
number of iterations required by standard iterative solvers tends to
decrease with increasing \Xavi{$Re$-numbers}. The predicted residual
scalings have been confirmed for all turbulent configurations
considered, \ie~homogeneous isotropic turbulence, Rayleigh–B\'{e}nard
convection, and bluff-body wakes, supporting the validity of the
proposed framework. In contrast, for the one-dimensional Burgers’
equation, the cost of solving the Poisson equation increases with
$Re$. This simplified model allows simulations over a much broader
range of Reynolds numbers, providing an extensive validation of the
theoretical scaling laws derived here.

\mbigskip

Overall, these findings indicate that, although the Poisson equation
remains the main bottleneck in incompressible CFD, its relative
computational cost may lessen for \Xavi{very} high $Re$-numbers. The
proposed theoretical framework thus provides a unified perspective on
how solver performance scales with $Re$-number and offers valuable
guidance for the development of next-generation preconditioning and MG
strategies for extreme-scale \Xavi{CFD}~simulations.

\section*{Acknowledgments}

\label{acks}

This work was financially supported by the SIMEX project
(PID2022-142174OB-I00) of {\it Ministerio de Ciencia e Innovaci\'{o}n}
MCIN/AEI/ 10.13039/501100011033 and the European Union Next
GenerationEU. \Adel{A.A.B. was financially supported by the EU's
  Horizon Europe programme under the Marie Sk\l odowska‑Curie grant
  agreement No.~101208388.  Views and opinions expressed are however
  those of the authors only and do not necessarily reflect those of
  the European Union or the European Research Executive
  Agency. Neither the European Union nor the granting authority can be
  held responsible for them.} Calculations \aft{corresponding to the
  simulations presented in Section~\ref{complex_flows}} were carried
out on the MareNostrum~5-GPP supercomputer at the BSC. \aft{The
  simulations presented in Section~\ref{HIT_tests} were performed
  using the Dutch national e-infrastructure with the support of the
  SURF Cooperative under grant no.~EINF-7119.} We thankfully
acknowledge these institutions.








\bibliographystyle{unsrt}


\bibliography{mybiblio} 



%
%

%



\end{document}